\documentclass[sigplan,9pt]{acmart}
\renewcommand\footnotetextcopyrightpermission[1]{}

\usepackage{textcomp}
\usepackage[font=small]{subcaption}
\usepackage{textgreek}
\usepackage{xspace}
\usepackage{pifont}
\usepackage{enumitem}
\usepackage{xcolor}
\usepackage{multirow}
\usepackage{algorithm}
\usepackage{algpseudocode}
\usepackage{framed}
\usepackage[locale=US]{siunitx}
\DeclareSIUnit{\microsecond}{\SIUnitSymbolMicro s}

\usepackage{tikz}
\usetikzlibrary{shapes.geometric}

\newcommand{\comp}[1]{%
  \tikz[baseline=-0.5ex]{%
    \node[circle, fill=black, inner sep=0pt, minimum size=1.8ex,
          text=white, font=\sffamily\bfseries\scriptsize] {#1};%
  }%
}

\definecolor{blinkblue}{HTML}{0072B2}
\definecolor{trtorange}{HTML}{D55E00}
\definecolor{vllmgreen}{HTML}{009E73}
\definecolor{sglangpurple}{HTML}{CC79A7}

\newcommand{\legendentry}[3]{%
  \tikz[baseline=-0.6ex]{%
    \draw[#1, line width=1.4pt] (0,0) -- (0.45,0);%
    \node[#1, #2, inner sep=1.1pt, line width=0.5pt] at (0.225,0) {};%
  }\,{\small #3}%
}

\newcommand{\figlegend}{%
  \par\vspace{4pt}%
  \centering%
  \legendentry{blinkblue}{circle, fill=blinkblue, draw=blinkblue}{Blink}\qquad%
  \legendentry{trtorange}{rectangle, fill=trtorange, draw=trtorange}{TRT-LLM}\qquad%
  \legendentry{vllmgreen}{regular polygon, regular polygon sides=3, fill=vllmgreen, draw=vllmgreen}{vLLM}\qquad%
  \legendentry{sglangpurple}{diamond, fill=sglangpurple, draw=sglangpurple}{SGLang}\qquad%
  \textbar\qquad%
  \tikz[baseline=-0.6ex]{\draw[black, line width=1.4pt] (0,0) -- (0.55,0);%
    \node[black, circle, fill=black, inner sep=0.8pt, line width=0.4pt] at (0.275,0) {};%
  }\,{\small Isolated}\qquad%
  \tikz[baseline=-0.6ex]{\draw[black, line width=1.2pt, dashed] (0,0) -- (0.55,0);%
    \node[draw=black, fill=white, circle, inner sep=0.8pt, line width=0.4pt] at (0.275,0) {};%
  }\,{\small Interference}%
  \par\vspace{2pt}%
}
\newcommand{\barlegendentry}[2]{%
  \tikz[baseline=-0.6ex]{%
    \node[rectangle, fill=#1, draw=#1, inner sep=0pt,
          minimum width=0.9em, minimum height=1.1ex] at (0.225,0) {};%
  }\,{\small #2}%
}
\newcommand{\barlegend}{%
  \par\vspace{3pt}%
  \centering%
  \barlegendentry{blinkblue}{Blink}\qquad%
  \barlegendentry{trtorange}{TRT-LLM}\qquad%
  \barlegendentry{vllmgreen}{vLLM}\qquad%
  \barlegendentry{sglangpurple}{SGLang}%
  \par\vspace{2pt}%
}

\newcommand{\figlegendnosuffix}{%
  \par\vspace{4pt}%
  \centering%
  \legendentry{blinkblue}{circle, fill=blinkblue, draw=blinkblue}{Blink}\qquad%
  \legendentry{trtorange}{rectangle, fill=trtorange, draw=trtorange}{TRT-LLM}\qquad%
  \legendentry{vllmgreen}{regular polygon, regular polygon sides=3, fill=vllmgreen, draw=vllmgreen}{vLLM}\qquad%
  \legendentry{sglangpurple}{diamond, fill=sglangpurple, draw=sglangpurple}{SGLang}%
  \par\vspace{2pt}%
}

\usepackage{tabularx}
\usepackage{booktabs}

\usepackage{pbalance}

\usepackage{chngcntr}

\usepackage{comment}  

\usepackage{placeins}

\usepackage[acronym, nohypertypes={acronym}]{glossaries}
\newacronym{CAT}{CAT}{Intel Cache Allocation Technology}
\newacronym{CLOS}{CLOS}{Class of Service}
\newacronym{dTLB}{dTLB}{Data Translation Look-aside Buffer}
\newacronym{DVFS}{DVFS}{Dynamic Voltage and Frequency Scaling}
\newacronym{IPC}{IPC}{Instructions per cycle}
\newacronym{ITL}{ITL}{Inter-Token Latency}
\newacronym{LLC}{LLC}{Last Level Cache}
\newacronym{LLM}{LLM}{Large Language Model}
\newacronym{MoE}{MoE}{Mixture-of-Experts}
\newacronym{RDT}{RDT}{Intel Resource Director Technology}
\newacronym{SLO}{SLO}{Service Level Objective}
\newacronym{SSE}{SSE}{Server-Sent Events}
\newacronym{SM}{SM}{Streaming Multiprocessor}
\newacronym{TLB}{TLB}{Translation Look-aside Buffer}
\newacronym{TPOT}{TPOT}{time-per-output-token}
\newacronym{TRT}{TRT}{TensorRT-LLM}
\newacronym{TTFT}{TTFT}{time-to-first-token}
\newacronym{AI}{AI}{Artificial Intelligence}
\newacronym{CUDA}{CUDA}{Compute Unified Device Architecture}

\usepackage{graphicx} 

\definecolor{takeawaybg}{gray}{0.92}
\newcommand{\takeaway}[2]{%
  \vspace{0.25em}
  \noindent
  \colorbox{takeawaybg}{\parbox{\dimexpr\linewidth-2\fboxsep-2\fboxrule\relax}{%
    \noindent\textit{\textbf{Takeaway:} #2}
  }}%
}

\makeatletter
\newcommand{\DeclareLatinAbbrev}[2]{%
  \DeclareRobustCommand{#1}{%
    \@ifnextchar{.}{\textit{#2}}{%
      \@ifnextchar{,}{\textit{#2.}}{%
        \@ifnextchar{!}{\textit{#2.}}{%
          \@ifnextchar{?}{\textit{#2.}}{%
            \@ifnextchar{)}{\textit{#2.}}{%
              {\textit{#2.,\ }}}}}}}}%
}
\makeatother
\DeclareLatinAbbrev{\eg}{e.g}
\DeclareLatinAbbrev{\Eg}{E.g}
\DeclareLatinAbbrev{\ie}{i.e}
\DeclareLatinAbbrev{\Ie}{I.e}
\DeclareLatinAbbrev{\etc}{etc}
\DeclareLatinAbbrev{\etal}{et~al}

\def\first {(\textit{i})\xspace}
\def\Second{(\textit{ii})\xspace}
\def\third {(\textit{iii})\xspace}

\newcommand{\systemname}[0]{\textsc{Blink}\xspace}

\newcommand{\smartparagraph}[1]{%
  \par\addvspace{0.6ex plus 0.1ex minus 0.05ex}%
  \noindent\textbf{#1}\xspace%
}
\AtBeginDocument{%
  }

\setcopyright{acmlicensed}
\copyrightyear{2018}
\acmYear{2018}
\acmDOI{XXXXXXX.XXXXXXX}
\acmConference[Conference acronym 'XX]{Make sure to enter the correct
  conference title from your rights confirmation email}{June 03--05,
  2018}{Woodstock, NY}
\acmISBN{978-1-4503-XXXX-X/2018/06}

\begin{document}

\title[Blink: CPU-Free LLM Inference by Delegating the Serving Stack to GPU and SmartNIC]{Blink: CPU-Free LLM Inference by Delegating the Serving Stack to GPU and SmartNIC}


\author{Mohammad Siavashi}
\affiliation{%
  \institution{KTH Royal Institute of Technology}
  \city{Stockholm}
  \state{}
  \country{Sweden}}
\email{}

\author{Mariano Scazzariello}
\affiliation{%
  \institution{RISE}
  \city{Stockholm}
  \state{}
  \country{Sweden}}
\email{}

\author{Gerald Q. Maguire Jr.}
\affiliation{%
  \institution{KTH Royal Institute of Technology}
  \city{Stockholm}
  \state{}
  \country{Sweden}}
\email{}

\author{Dejan Kosti\'c}
\affiliation{%
  \institution{KTH Royal Institute of Technology}
  \city{Stockholm}
  \state{}
  \country{Sweden}}
\email{}

\author{Marco Chiesa}
\affiliation{%
  \institution{KTH Royal Institute of Technology}
  \city{Stockholm}
  \state{}
  \country{Sweden}}
\email{}

\renewcommand{\shortauthors}{Siavashi et al.}

\begin{abstract}

\gls{LLM} inference is rapidly becoming a core datacenter service, yet current serving stacks keep the host CPU on the critical path for orchestration and token-level control. This makes LLM performance sensitive to CPU interference, undermining application colocation and forcing operators to reserve CPU headroom, leaving substantial capacity unutilized.
We introduce \systemname, an end-to-end serving architecture that \textit{removes} the host CPU from the steady-state inference path by redistributing responsibilities across a SmartNIC and a GPU. \systemname offloads request handling to the SmartNIC, which delivers inputs directly into GPU memory via RDMA, and replaces host-driven scheduling with a \textit{persistent} GPU kernel that performs batching, scheduling, and KV-cache management \textit{without} CPU involvement.
Evaluated against TensorRT-LLM, vLLM, and SGLang, \systemname outperforms all baselines even in isolation, reducing pre-saturation P99 \gls{TTFT} by up to \num{8.47}$\times$ and P99 \gls{TPOT} by up to \num{3.40}$\times$, improving decode throughput by up to \num{2.1}$\times$, and reducing energy per token by up to \qty{48.6}{\percent}. Under CPU interference, \systemname maintains stable performance, while existing systems degrade by up to two orders of magnitude.

\end{abstract}

%
%



\maketitle
\fancyhead{}
\glsresetall
\section{Introduction}
\begin{figure}[t]
    \centering
    \includegraphics[width=\columnwidth]{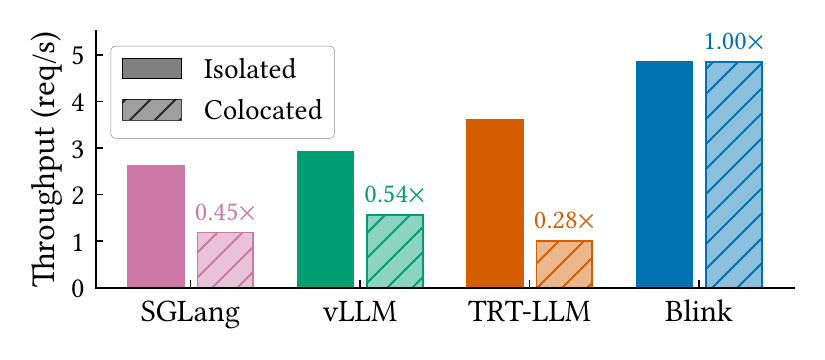}
    \caption{Achieved throughput of 4 LLM serving systems on \mbox{Qwen-3} 30B-A3B (MoE) at 4\,req/s offered load, under isolated and colocated execution. \systemname is unaffected by colocation while baselines degrade. Annotations show the colocated-to-isolated throughput ratio.}
    \label{fig:intro-teaser}
\end{figure}

\glspl{LLM} are a foundational technology for modern \gls{AI} systems, powering search, translation, summarization, assistants, coding, and enterprise automation. As deployments scale to large user bases~\cite{chatgpt_users}, including latency-sensitive applications~\cite{fastserve,distserve}, inference cost and complexity are becoming primary concerns for datacenter operators~\cite{hygen,ConServe}. Although inference pipelines rely heavily on GPUs for tensor computation, standard architectures still center the server CPU for request handling, orchestration, and data movement~\cite{orca,vllm,mlsys_scheduling}.

\smartparagraph{LLM serving systems are increasingly CPU-dependent.} As quality-of-service demands become stricter, the control and data planes running on the server are growing increasingly complex. For example, after Orca~\cite{orca} introduced iteration-level scheduling, many systems adopted this approach, which requires handing control back to the CPU of the server after every token (or every few tokens). Moreover, recent control-plane optimizations such as chunked prefill~\cite{sarathi_serve}, which splits long prompts into smaller chunks and interleaves them with decode tokens to reduce latency stalls, further complicate CPU-side logic. Although these techniques improve LLM serving performance, they also add significant CPU-side overheads, making inference susceptible to interference from colocated workloads~\cite{shenango}.


\smartparagraph{The impact of CPU interference.} The dominant LLM deployment model dedicates an entire machine to a single workload, simplifying \gls{SLO} management and predictability but often leaves CPU resources stranded~\cite{hygen,ConServe,Niyama}. 
Reclaiming underused CPU resources on machines running GPU-intensive LLM workloads, \textit{without} degrading performance, would allow datacenters and cloud providers to generate additional revenue from the hundreds of thousands of machines currently dedicated to LLM inference.
%
%
We quantify this fragility in Figure~\ref{fig:intro-teaser}: under colocation with CPU-intensive workloads, three production serving systems (SGLang~\cite{sglang}, vLLM~\cite{vllm}, TRT-LLM~\cite{tensorrt_llm}) retain only 28--54\% of their isolated throughput, highlighting the impact of CPU interference on inference performance.

\smartparagraph{Existing SmartNIC offloads primarily target the data path.}
LLM inference consists of two tightly coupled components. The \textit{data path} carries requests and model data between the network and the GPU, including packet processing, tensor transfers, and I/O movement. The \textit{control path} governs how inference proceeds, including batching, scheduling, accelerator coordination, and state management during per-token decoding. 
SmartNICs are a natural target for offloading such functionality, as they lie on the critical path of request handling between the network and the GPU and bypassing the kernel.
Prior SmartNIC-based systems successfully offload data-path tasks such as packet parsing, deep packet inspection, storage-client logic, and request handling to SmartNICs~\cite{Kfoury_2024,os2g,10.1145/3589974,298581}, improving throughput for conventional one-shot inference workloads such as traditional Deep Neural Network (DNN) execution. In these workloads, each request executes as a largely self-contained forward pass, with scheduling decisions occurring at coarse granularity rather than at every token, resulting in a relatively lightweight control path with limited CPU involvement.

\smartparagraph{Autoregressive LLMs shift the bottleneck to the control path.}
LLM serving differs fundamentally. Autoregressive decoding transforms inference into a long-lived, stateful process in which each generated token depends on previously produced state. Latency-sensitive operations such as KV-cache management, batching decisions, and token streaming are tightly coupled to per-token scheduling. As a result, the control path becomes part of the critical loop.
Existing SmartNIC-based inference systems offload portions of request handling or data movement~\cite{10.1145/3589974,298581}, but they do not address autoregressive decoding. Token-by-token execution, placement, and flow control repeatedly interact with GPU-resident state, while scheduling and coordination remain CPU-centric. Consequently, the critical loop remains anchored in host-managed mechanisms, leaving the server CPU on the steady-state inference path.
This shift makes removing the CPU fundamentally harder, as both control and data responsibilities must be restructured across the system. This fundamentally \textit{challenges the traditional separation between data and control paths}: decisions that were previously coarse-grained and host-driven must now be made at token-level granularity, tightly coupled with GPU state.

\smartparagraph{Why CPU-free inference is hard.}
Achieving a CPU-free\footnote{CPU-free refers to removing the host CPU from the inference critical path.} inference stack that delivers production-grade performance and predictability raises several challenges.

First, \textit{eliminating host-driven scheduling and control} is fundamentally difficult. Autoregressive decoding turns inference into a long-lived, stateful process with fine-grained, per-token decisions (\eg batching, KV-cache management, and token generation). These decisions must be executed without host intervention, while ensuring forward progress and fairness, and without allowing control logic to interfere with GPU resources needed for inference.

Second, the system must \textit{coordinate thousands of concurrent requests} without a host scheduler. This requires managing queuing, backpressure, and request lifecycles (including early exits and failures) across mismatched timescales (\ie microsecond network events, millisecond GPU kernels, and second-scale request lifetimes) using distributed control across the NIC and GPU.

Third, removing the host CPU requires rethinking \textit{transport and request handling}. Conventional TCP/IP stacks and host drivers are tightly coupled with system memory and the OS scheduler, making them difficult to bypass without redesigning the data plane, especially under the limited compute budget of NIC-resident processors.

Finally, enabling \textit{efficient network-to-GPU} communication demands direct data movement. This requires integrating RDMA and peer-to-peer DMA (\eg GPUDirect RDMA) into the serving stack to eliminate redundant copies and CPU involvement, without sacrificing throughput, latency, or fault isolation.

\smartparagraph{\systemname: removing the server's main CPU from the critical path.}
In this paper, we present \systemname, a novel LLM inference system that addresses these challenges by removing the host CPU from the steady-state inference path and redistributing responsibilities across a GPU and a SmartNIC DPU.
To \textit{eliminate host-driven scheduling and control}, \systemname replaces the traditional decode loop with CUDA persistent kernels that implement a GPU-resident control plane, handling continuous batching, iteration-level scheduling, and Paged KV-cache management without per-token host interaction.
To \textit{coordinate thousands of concurrent requests} without a central scheduler, this control plane operates over shared data structures, enabling fairness, backpressure, and forward progress entirely on the GPU.
To \textit{remove dependence on host-based transport} and request handling, \systemname offloads protocol parsing and request admission to the SmartNIC, bypassing kernel-mediated interfaces.
Finally, to \textit{enable efficient data movement}, the system uses one-sided RDMA and shared ring buffers to move data directly between the network and GPU VRAM, eliminating copies through host memory and CPU mediation while maintaining high throughput and low latency.

\smartparagraph{Evaluation benefits.}
We compare \systemname with three state-of-the-art production LLM serving systems (TensorRT-LLM (TRT), vLLM, and SGLang) across four models under both isolated and multi-tenant execution.
Even in \textit{isolated settings}, \systemname outperforms all baselines, reducing pre-saturation P99 time-to-first-token (TTFT) by up to 8.47$\times$ and P99 time-per-output-token (TPOT) by up to 3.40$\times$, and improving decode throughput by up to 2.1$\times$. These gains stem from eliminating per-token host interaction via a GPU-resident control loop.
Under \textit{host multi-tenancy interference}, \systemname’s latency and throughput remain stable (within experimental variance of isolated values), whereas all baselines degrade by one to two orders of magnitude. As a result, \systemname sustains up to 6.46$\times$ higher decode throughput (tokens/s) and up to 4.87$\times$ higher request throughput (requests/s) under CPU contention.
These improvements translate into up to 48.6\% lower energy per token in isolation, and up to 70.7\% under interference.
%
We plan to open-source \systemname upon publication.

\smartparagraph{Contributions.} This paper makes four primary contributions:
\begin{itemize}[leftmargin=*,noitemsep]
\item \textit{Identifying control-path dependence as the key bottleneck in LLM inference.}
We show that modern LLM serving systems rely on fine-grained, host-driven control, making them sensitive to CPU interference and limiting efficient resource sharing.

\item \textit{A CPU-free LLM inference architecture.}
We introduce \systemname, a new class of serving systems that removes the host CPU from the steady-state critical path by redistributing control and data responsibilities across a GPU and a SmartNIC DPU.

\item \textit{GPU-resident control via persistent kernels.}
We design a GPU-resident scheduler that replaces the host-driven decode loop, enabling continuous batching, scheduling, and KV-cache management without per-token CPU interaction.

\item \textit{Strong performance and isolation benefits.}
\systemname outperforms state-of-the-art systems even in isolation (up to 2.1$\times$ higher throughput and 8.47$\times$ lower pre-saturation P99 TTFT), while maintaining stable performance under CPU interference where existing systems degrade by orders of magnitude.
\end{itemize}

\noindent

\section{Background and Motivation}
\label{sec:motivation}

\subsection{The Host CPU as a Latency Bottleneck}
\label{sec:host-bottleneck}

In every mainstream \gls{LLM} serving stack, the host CPU orchestrates each iteration of autoregressive decoding. Request admission, continuous batching, KV-cache block management, and CUDA kernel dispatch all execute on host threads. Even with CUDA Graphs~\cite{cuda_graphs} amortizing individual kernel launch costs, the scheduler must return to the host after every decode step to update batch membership, manage the KV-cache block table, and dispatch the next graph.
For instance, a single vLLM instance uses a multi-process architecture with an API server, an engine core, and one GPU worker per GPU~\cite{vllm_proc_arch_v016}; this footprint scales with the chosen parallelism strategy, causing the number of host processes to grow rapidly with the deployment configuration~\cite{vllm_proc_count_v016}.


This tight host-device coupling means that any perturbation to the CPU, \ie preemption, cache eviction, or elevated page-walk latency, directly inflates the \gls{ITL}. Additionally, the GPU idles while the host rebuilds the microarchitectural state, and because decoding is iterative, the penalty compounds across hundreds of output tokens. This feedback loop between host jitter and token-level latency is quantified in \S\ref{sec:pmc}.

\smartparagraph{Recent optimizations reduce but do not eliminate host\linebreak[4] involvement.} The severity of host-side scheduling overhead has not gone unnoticed. Profiling studies report that CPU scheduling can consume up to \qty{50}{\percent} of end-to-end inference latency on fast accelerators~\cite{mlsys_scheduling}. In response, recent engine revisions (\eg vLLM's V1 engine architecture~\cite{vllm_v1} and SGLang's overlapped scheduling~\cite{sglang_v04}) pipeline host-side work with GPU execution, reducing the scheduling tax under normal operating conditions. These optimizations mitigate CPU involvement but they do not remove the CPU from the critical path. Under colocation, co-tenants evict shared LLC lines and TLB entries, inflating the CPU operations that the overlap is designed to hide. Once host-side work exceeds the GPU execution interval available to mask it, the excess latency surfaces directly in each token's generation time. As we quantify in \S\ref{sec:pmc}, even moderate interference is sufficient to break this overlap, because any CPU work that remains on the critical path turns shared microarchitectural contention into per-token latency.

\smartparagraph{Colocation is the norm, not the exception.} Production clusters aggressively pack workloads to chase high utilization~\cite{shenango}. Cluster schedulers routinely colocate latency-critical services with best-effort or batch jobs, relying on priorities, cgroups, and NUMA partitioning to manage interference. Operators have developed a mature toolkit for isolation (\gls{LLC} partitioning (Intel CAT), huge pages, and \gls{DVFS} tuning), yet all of these mitigations assume that the latency-sensitive workload can tolerate \emph{some} CPU involvement. For LLM inference, where the CPU participates in every token, even well-managed colocation introduces \emph{per-token jitter} that accumulates into \gls{SLO} violations at the tail.

\subsection{Motivating Measurement: CPU Interference}
\label{sec:pmc}
We begin with a simple experiment to illustrate the impact of CPU interference before analyzing root causes in~\S\ref{sec:characterization}.
We conduct controlled experiments using vLLM (v0.13)~\cite{vllm} serving Llama-3 8B~\cite{llama3} on an NVIDIA H100 GPU driven by a dual-socket Xeon Gold server (for details, see Table~\ref{tab:system-config}), with ShareGPT v3~\cite{sharegpt_v3} conversation traces (mean input/output lengths of \num{1019}/\num{463} tokens). CUDA Graphs are enabled in all configurations; hyper-threading and \gls{DVFS} are disabled following standard practice for microarchitectural measurement. As the colocated interferer, we run \texttt{pbzip2}~\cite{pbzip2} compressing a large file, measuring both application-level \gls{SLO} metrics and hardware performance counters via \texttt{perf stat}. The serving engine is fully warmed up before measurement begins; profiling spans the entire run including ramp-up and drain, which is conservative as lightly loaded transients dilute the steady-state interference effect.

Table~\ref{tab:colocation_summary} reveals the severity of the problem. Under interference from \texttt{pbzip2}, throughput drops by \num{3.8}$\times$, while P99 \gls{TTFT} inflates by up to \num{139}$\times$. Even moderate interference causes significant degradation, indicating that LLM serving performance is highly sensitive to CPU contention.
This degradation is accompanied by substantial increases in cache misses (LLC), address translation misses (dTLB), and memory stalls, consistent with contention in shared CPU resources.
A natural question is whether this fragility can be mitigated using standard datacenter techniques; \S\ref{sec:characterization} shows that such approaches are insufficient to eliminate the underlying CPU bottleneck.


\sisetup{
  group-separator = {\,},
  group-minimum-digits = 4,
  table-number-alignment = center
}

\sisetup{
  group-separator = {\,},
  group-minimum-digits = 4
}

\begin{table}[t]
\caption{Impact of colocation on vLLM serving latency and microarchitectural counters. H100, Llama-3 8B, 
\qty{7}{req/s}, w\slash~CUDA Graphs.}
\label{tab:colocation_summary}
\centering
\resizebox{0.85\linewidth}{!}{%
\begin{tabular}{lccc}
{} & \textbf{Baseline} & \multicolumn{2}{c}{\textbf{Interference}} \\
{} & {} & \textbf{12$\times$} & \textbf{24$\times$} \\
\midrule
Throughput (tok/s)         & \num{7475}  & \num{4554}  & \num{1961}  \\
Mean \gls{TTFT} (ms)       & \num{73.7}  & \num{4865}  & \num{16552} \\
P99 \gls{TTFT} (ms)        & \num{150}   & \num{6366}  & \num{20959} \\
Mean TPOT (ms)       & \num{13.0}  & \num{13.6}  & \num{14.8}  \\
P99 TPOT (ms)        & \num{14.4}  & \num{18.0}  & \num{32.1}  \\
P99 \gls{ITL} (ms)         & \num{67.9}  & \num{110.6} & \num{176.8} \\
\midrule
\gls{IPC}                  & \num{1.53}  & \num{1.08}  & \num{0.72}  \\
\gls{LLC} miss rate (\%)   & \num{7.0}   & \num{43.2}  & \num{71.6}  \\
LLC stall cycles        & \num{450}\,M   & \num{2586}\,M  & \num{5037}\,M  \\
dTLB load misses & \num{6}\,M     & \num{8}\,M     & \num{10}\,M    \\
\texttt{walk\_active}  & \num{383}\,M   & \num{920}\,M   & \num{1454}\,M  \\
CPU migrations             & \num{6}     & \num{20}    & \num{27}    \\
\end{tabular}
}
\end{table}


\section{Limitations of CPU-Based Mitigations}
\label{sec:characterization}
We evaluate whether standard datacenter mitigations, such as huge pages, cache partitioning, CPU pinning, or dynamic core reallocation, can address
the performance degradation observed in~\S\ref{sec:pmc}.
In~\S\ref{sec:microarch}, we trace the root cause of the performance degradation using hardware performance counters and systematically evaluate each mitigation. In~\S\ref{sec:tuning-limits}, we show that OS tuning is not a solution. All paths converge on the same irreducible overhead: the overhead of CPU-mediated orchestration itself. This finding motivates our new alternative architecture that entirely removes the host CPU from the steady-state inference loop (\S\ref{sec:implications}).

\subsection{Microarchitectural Root Causes}
\label{sec:microarch}

\smartparagraph{Address translation intensifies \gls{LLC} contention.} Breaking down the LLC stall budget (from Table~\ref{tab:colocation_summary}) exposes a two-stage amplification loop driven by address translation. In the baseline case, page-walk activity (\ie page-table traversal on a TLB miss, such as dTLB load misses and \texttt{walk\_active}) accounts for \qty{85}{\percent} of all LLC stall cycles, consistent with the translation-heavy behavior of Python-based serving stacks operating over fragmented virtual address spaces~\cite{10.1007/978-3-031-15074-6_14}. When interference is introduced, the total number of \gls{TLB} misses rises only moderately (\num{1.6}$\times$), but each miss becomes substantially more costly: the interferer's frequent memory-management operations (\eg \texttt{madvise}, \texttt{mprotect}, and \texttt{munmap}) induce \gls{TLB} invalidations that force re-translation and concurrently contaminate the shared \gls{LLC} with additional data and page-table metadata. Page walks that previously hit page-table entries residing in the \gls{LLC} now must proceed all the way to DRAM, driving up \texttt{walk\_active} cycles by \num{3.8}$\times$. 
In aggregate, LLC stall cycles increase by \num{11.2}$\times$, while the fraction of stalls attributed to page walks falls to under \qty{25}{\percent}, not because page walks become cheaper, but because \gls{LLC} data-access misses escalate even more sharply. This two-level amplification, where TLB invalidations force page walks into an already-polluted LLC and drive up data-access misses, creates a cross-address-space interference mechanism that standard per-process mitigations cannot resolve.

\smartparagraph{Tail latency exhibits high variability.} 
Even in isolation, host-mediated orchestration introduces jitter: individual token latencies spike due to batch scheduling, KV-cache management, and CUDA dispatch variance, even when per-request averages stay low. This is visible in the gap between P99 \gls{ITL} and P99 \gls{TPOT} (\qty{67.9}{\milli\second} vs. \qty{14.4}{\milli\second}, a 4.7$\times$ difference). Under interference, this jitter is amplified further: inter-kernel dispatch gaps widen from $\approx$\qty{1}{\microsecond} to \qty{40}{\microsecond} (measured via NVIDIA Nsight Systems~\cite{nvidia_nsight}), compounding across hundreds of output tokens per request.

\takeaway{}{CPU interference degrades LLM serving severely: TLB invalidations and LLC pollution share the same microarchitectural resources, amplifying each other's impact.}

\vspace{-1em}
\subsection{OS Tuning Is Insufficient}
\label{sec:tuning-limits}
A natural question is whether standard operating system and hardware mitigations (\eg larger pages, cache partitioning, CPU affinity, and scheduling priorities) can restore performance under colocation. We use each mitigation as a controlled variable to systematically isolate the root cause of the degradation. The results reveal that interference enters through multiple microarchitectural channels, but the dominant bottleneck is not any single channel---it is CPU involvement on the critical path itself.

Unless otherwise noted, the experiments in this subsection use the same \gls{LLM} server, the same H100 server, and CUDA Graphs enabled as described in \S\ref{sec:pmc}, but with the interferer limited to 24 threads and \num{128} requests at \qty{7}{req/s}. Additionally, we use a synthetic workload (random input \& output lengths of \qty{1024} \& \qty{512} tokens) to maximise batch occupancy and stress the host scheduling path.

\smartparagraph{Huge pages (victim and interferer): negligible effect.} 
Larger pages reduce TLB pressure by covering more memory per entry, potentially reducing the page-walk overhead identified in \S~\ref{sec:microarch}. Table~\ref{tab:page_ablation} shows that neither 2\,MB pages for vLLM nor 1\,GB gigantic pages for the interferer restore isolation-level performance. With 2\,MB pages, all latency and throughput metrics remain within measurement noise; \gls{dTLB} load misses drop only \qty{16}{\percent}, consistent with the limited \gls{TLB} reach benefit of 2\,MB pages for Python-dominated working sets whose host-side footprint is dominated by small objects and metadata. Allocating the interferer's working set from a 1\,GB hugepage pool does not help: P99 \gls{ITL} \emph{worsens} and \gls{LLC} miss rates are unchanged, because gigantic pages reduce the interferer's own \gls{TLB} misses but not its \gls{LLC} footprint, similarly to what we observed in \S\ref{sec:microarch}.

\sisetup{
  group-separator = {\,},
  group-minimum-digits = 4
}

\begin{table}[t]
\centering
\small
\caption{Ablation: page size and gigantic pages do not restore isolation (H100, CUDA Graphs enabled, 128 requests at \qty{7}{req/s}). All runs include interference. Isolation baseline: \qty{7697}{tok/s} throughput, \qty{13.5}{\milli\second} mean \gls{TPOT}, \qty{5.9}{\percent} \gls{LLC} miss rate.}
\label{tab:page_ablation}
\resizebox{.95\linewidth}{!}{%
\begin{tabular}{lccc}
 & \textbf{4\,KB pages} & \textbf{2\,MB pages} & \textbf{1\,GB (interferer)} \\
\midrule
Throughput (tok/s) & \num{4813} & \num{5053} & \numrange{4540}{5298} \\
\midrule
P50 \gls{TTFT} (ms) & \num{12276} & \num{12612} & \numrange{12578}{15205} \\
P99 \gls{TTFT} (ms) & \num{29208} & \num{26780} & \numrange{29191}{32270} \\
\midrule
P50 \gls{TPOT} (ms) & \num{19.8} & \num{19.9} & \numrange{17.9}{19.6} \\
P99 \gls{TPOT} (ms) & \num{25.0} & \num{25.4} & \numrange{27.2}{30.5} \\
P99 \gls{ITL} (ms)  & \num{70.1} & \num{72.2} & \numrange{73.2}{84.8} \\
\midrule
\gls{LLC} miss rate (\%) & \num{71.3} & \num{71.2} & \numrange{68.5}{70.3} \\
\gls{dTLB} load misses   & \num{8.8}\,M & \num{7.4}\,M & \numrange{7.8}{10.0}\,M \\
\texttt{walk\_active}    & \num{1132}\,M & \num{985}\,M & \numrange{779}{970}\,M \\
\end{tabular}
}
\end{table}

\smartparagraph{Core pinning: effective but impractical at scale.} Dedicating physical cores with core pinning eliminates CPU migrations and prevents direct preemption by co-tenants, making it the strongest single-mechanism mitigation we evaluate. Following NVIDIA's Certified Systems Configuration Guide mandate of a minimum of six physical cores per GPU~\cite{nvidia_cert_guide}, we pin vLLM to cores 0--5 and confine the \texttt{pbzip2} interferer to the remaining cores. To stress-test pinning under realistic conditions, this experiment uses production-representative traffic: ShareGPT conversations with Poisson arrivals at \qty{12}{req/s} measured over a \qty{60}{\second} window (unlike the micro-benchmarks above, which maximise batch occupancy). Table~\ref{tab:pinning_ablation} shows the results.

\begin{table}[t]
\centering
\small
\caption{Core pinning (6 dedicated cores) under \texttt{pbzip2} interference. ShareGPT workload, Poisson arrivals at \qty{12}{req/s}, \qty{60}{\second} window.}
\label{tab:pinning_ablation}
\resizebox{.95\linewidth}{!}{%
\begin{tabular}{lccc}
 & \textbf{Isolation} & \textbf{Interference} & \textbf{$\Delta$\%} \\
\midrule
Total completed requests           & \num{502}      & \num{415}      & -17.3\,\% \\
Mean throughput (tok/s)            & \num{4130.50}  & \num{3457.85}  & -16.3\,\% \\
Mean throughput (req/s)            & \num{8.37}     & \num{6.92}     & -17.3\,\% \\
\midrule
P50 \gls{TTFT} (ms)                & \num{10052.08} & \num{12539.18} & +24.7\,\% \\
P99 \gls{TTFT} (ms)                & \num{22409.94} & \num{23968.67} & +7.0\,\% \\
P99.9 \gls{TTFT} (ms)              & \num{23925.64} & \num{25732.75} & +7.6\,\% \\
\midrule
P50 \gls{TPOT} (ms)                & \num{29.42}    & \num{37.90}    & +28.8\,\% \\
P99 \gls{TPOT} (ms)                & \num{162.57}   & \num{192.45}   & +18.4\,\% \\
P99.9 \gls{TPOT} (ms)              & \num{499.64}   & \num{641.10}   & +28.3\,\% \\
\midrule
P50 \gls{ITL} (ms)                 & \num{10.56}    & \num{12.87}    & +21.9\,\% \\
P99 \gls{ITL} (ms)                 & \num{13.83}    & \num{16.48}    & +19.2\,\% \\
P99.9 \gls{ITL} (ms)               & \num{16.45}    & \num{21.44}    & +30.3\,\% \\
\midrule
Mean prefill throughput (tok/s)    & \num{1375.08}  & \num{1223.27}  & -11.0\,\% \\
Median prefill throughput (tok/s)     & \num{161.87}   & \num{116.05}   & -28.3\,\% \\
\midrule
Mean decode throughput (tok/s)     & \num{2780.60}  & \num{2274.77}  & -18.2\,\% \\
Median decode throughput (tok/s)      & \num{2189.09}  & \num{1778.00}  & -18.8\,\% \\
\end{tabular}
}
\end{table}

We pin vLLM to six physical cores on the GPU-local socket (\ie 6 of 24 cores per socket, or \qty{25}{\percent}). Even then, interference causes significant degradation across all metrics (Table~\ref{tab:pinning_ablation}): tail latency inflates by up to \qty{30.3}{\percent} (P99.9 \gls{ITL} and \gls{TPOT}), decode throughput drops by \qty{18.2}{\percent}, and the server completes \qty{17.3}{\percent} fewer requests in the same window. The reason is structural: pinning removes scheduler contention, but the LLC, memory bandwidth, and socket interconnect remain shared with the co-located workload.
%
Moreover, dedicating a minimum of six physical CPU cores per GPU~\cite{nvidia_cert_guide} for a dense 8$\times$\,H100 node, requires 48 dedicated physical cores---consuming all cores on our dual-socket Xeon server (24 cores per socket) and \qty{38}{\percent} of even a high-end dual-socket EPYC platform (64 cores per socket). These dedicated cores sit largely idle between decode iterations, yet cannot be reclaimed for other tenants without re-introducing scheduling interference. In effect, core pinning converts a shared machine back into a dedicated one~\cite{shenango}, eliminating the economic rationale for colocation; thus, core pinning is impractical at scale.

\smartparagraph{Hardware cache partitioning: eliminates LLC contention but not the latency overheads.} \gls{CAT} requires core pinning as a prerequisite (a \gls{CLOS} is assigned to the \emph{core}, not the process, and is overridden on context switches), so we evaluate whether adding cache partitioning provides additional benefit on top of pinning. We incrementally allocate from 1 to 12 \gls{LLC} cache ways to vLLM's pinned cores under interference. Table~\ref{tab:cat_ablation} summarizes the results.

\begin{table}[h]
\centering
\caption{Effect of \gls{CAT} cache-way allocation under interference (H100, CUDA Graphs, \num{128} requests at \qty{7}{req/s}, dedicated CPU cores). \gls{LLC} miss rate drops $8.5\times$, yet P99 \gls{ITL} is virtually unchanged.}
\label{tab:cat_ablation}
\resizebox{.9\linewidth}{!}{%
\begin{tabular}{lccccc}
Cache ways & \num{1} & \num{3} & \num{5} & \num{7} & \num{12} \\
\midrule
\gls{LLC} miss rate (\%) & \num{57.6} & \num{26.6} & \num{11.1} & \num{7.0} & \num{6.8} \\
\gls{IPC} & \num{1.16} & \num{1.31} & \num{1.48} & \num{1.52} & \num{1.55} \\
LLC stall cycles & \num{3169}\,M & \num{1750}\,M & \num{741}\,M & \num{428}\,M & \num{442}\,M \\
\gls{dTLB} load misses & \num{7.4}\,M & \num{7.1}\,M & \num{7.1}\,M & \num{7.0}\,M & \num{7.0}\,M \\
\texttt{walk\_active} & \num{895}\,M & \num{596}\,M & \num{462}\,M & \num{420}\,M & \num{400}\,M \\
\midrule
P99 \gls{TTFT} (ms) & \num{29675} & \num{26627} & \num{27146} & \num{26352} & \num{26157} \\
P99 \gls{TPOT} (ms) & \num{23.3} & \num{22.6} & \num{21.3} & \num{20.9} & \num{21.3} \\
P99 \gls{ITL} (ms) & \num{55.6} & \num{55.6} & \num{53.4} & \num{54.4} & \num{54.0} \\
\end{tabular}
}
\end{table}

Allocating enough cache ways to vLLM recovers the \gls{LLC} miss rate to near-isolated levels and significantly reduces \gls{LLC} stall cycles. The mechanism is indirect: \gls{CAT} does not partition the \gls{TLB}, so the number of \gls{dTLB} misses stays constant across all configurations ($\approx$\qty{7}M). However, when a \gls{TLB} miss occurs, the CPU must walk the page table to resolve the physical address, and those page-table entries themselves reside in the \gls{LLC}. With more cache ways allocated to vLLM, page-table entries are more likely to remain \gls{LLC}-resident, making each walk shorter.At 7 ways, the \gls{LLC} miss rate reaches \qty{7.0}{\percent} (vs.\ \qty{6.8}{\percent} with all 12 ways) and \gls{LLC} stall cycles drop 7.4$\times$
7.4$\times$. 

Despite fully eliminating \gls{LLC} contention, \textit{tail latency is virtually unchanged}: P99 \gls{ITL} ranges from \qtyrange{53.4}{55.6}{\milli\second} across all configurations, a spread of less than \qty{4}{\percent}. Hardware cache partitioning does not improve latencies because the dominant overhead, \textit{i.e.}, host-side scheduling jitter and \gls{CUDA} dispatch, \textit{i.e.}, is unaffected by cache allocation (see \S\ref{sec:implications}.

Moreover, achieving this \gls{LLC} recovery comes at a steep cost to co-tenants. Our server has 12 \gls{LLC} ways; dedicating 7 to vLLM leaves only 5 ways for all remaining workloads on the socket, throttling their cache capacity by \num{2.4}$\times$. In a multi-GPU node, each GPU's serving process would need its own 7-way partition, quickly exhausting the available ways, making the mechanism simply unscalable. We also note that scheduling priority (\texttt{nice\,-20}) likewise has no measurable effect.

\takeaway{}{Even fully eliminating LLC contention does not restore tail latency: host-side scheduling jitter and CUDA dispatch overhead remain on the critical path regardless of cache allocation.}
\vspace{.03in}


\smartparagraph{Practical limitations of RDT.}
\gls{RDT}~\cite{intel_rdt}, of which \gls{CAT} is a component, is also impractical as an operator-level isolation mechanism: \gls{CLOS} entries are scarce (9--15 per CPU) and must be contiguous; \gls{CLOS} assignments are not preserved across context switches --- when another process is scheduled on the same core, its \gls{CLOS} overrides the previous one --- so maintaining isolation requires dedicated cores or kernel-level modifications to save and restore \gls{CLOS} state~\cite{phoenix}; \gls{CAT} addresses only cache occupancy, not memory bandwidth; and \gls{RDT} is Intel-specific, with no equivalent on AMD or ARM platforms. It is a low-level mechanism, not a turnkey isolation toolkit.


\smartparagraph{Host-side orchestration jitter is directly observable.} To confirm that host-side overhead is the residual bottleneck, we profile a single transformer layer's \texttt{forward()} with PyTorch~\cite{pytorch}. Under interference, every host-side operation inflates: attention dispatch by \qty{104}{\percent}, \texttt{cudaLaunchKernel} by \qty{115}{\percent}, KV-cache dispatch (\texttt{reshape\_and\_cache\_flash}) by \qty{172}{\percent}, and metadata operations such as \texttt{aten::empty} by \qty{81}{\percent}. Crucially, GPU-side kernel execution times are unchanged (\eg matmul remains at \qtyrange{0.41}{0.42}{\milli\second} per call), confirming that the accelerator is simply \emph{starved} by its host.

\subsection{Implications: The Case for Host Removal}
\label{sec:implications}

Our analysis reveals a fundamental architectural mismatch: today's serving systems place a fragile, interference-sensitive CPU on the critical path of every token, in an environment where interference is pervasive. Every mitigation we tested
either had negligible effect or left a persistent performance residual.
These are \emph{not} implementation bugs; they reflect a structural mismatch between the iterative host--device coupling of current serving stacks and the shared-resource reality of multi-tenant datacenters.

\smartparagraph{Resource partitioning does not resolve the architectural mismatch.} An operator might combine all preceding mitigations: dedicate physical cores via \texttt{cpuset}, partition \gls{LLC} ways via \gls{CAT}, and disable hyperthreading \ie approximating hardware-level socket partitioning---but at a prohibitive cost. Dynamic core reallocation systems such as Caladan~\cite{caladan} and Shenango~\cite{shenango} adjust core assignments at microsecond granularity, but optimize \emph{how much CPU capacity} a tenant receives, not \emph{whether the CPU is on the critical path}. \gls{LLC} and memory-bus bandwidth are socket-wide resources that per-core reallocation does not effectively partition, and both systems require applications to link against custom userspace threading and networking runtimes in place of standard POSIX APIs---a requirement incompatible with the Python/PyTorch stacks underlying all major \gls{LLM} serving engines.

\takeaway{}{
Achieving interference immunity requires removing the host CPU entirely from the steady-state inference loop.}

\section{\systemname Design}
\label{sec:design}
\subsection{Overview}
\label{sec:designOverview}

The design principle behind \systemname is to make the host CPU a \emph{provisioning plane} rather than a \emph{data plane}: the host loads the model and captures CUDA graphs once at startup, then exits the inference path entirely.
\systemname is an end-to-end LLM inference serving system with five core goals:
(1)~\textbf{high concurrency with predictable latency};
(2)~\textbf{CPU-free steady state} --- after one-time initialization, host CPU is off the inference path;
(3)~\textbf{zero-copy prompt and token movement} between the network fabric and GPU memory via one-sided RDMA;
(4)~\textbf{compatibility with existing models} without requiring any model modification; and
(5)~\textbf{API compatibility} with OpenAI-style HTTP endpoints and \gls{SSE} streaming semantics, enabling drop-in deployment.

\begin{figure}[t]
    \centering
    \includegraphics[clip, width=\columnwidth]{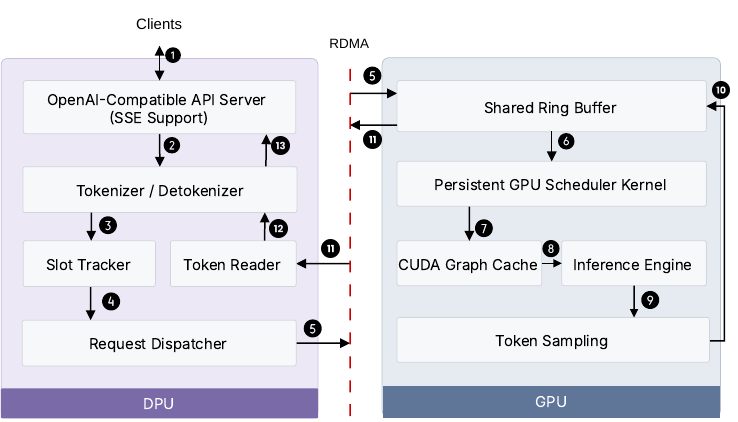}
    \caption[Architecture of Blink.]{Architecture of \systemname. After one-time initialization, the host CPU is idle; the steady-state inference path involves only the GPU and the DPU. Filled circles 
    trace the lifecycle of a single request.}
    \label{fig:axon-high-level-architecture}
\end{figure}

Fig.~\ref{fig:axon-high-level-architecture} shows the high-level architecture and traces the lifecycle of prompts through the inference path (\comp{1}--\comp{13}).
\systemname comprises two runtime planes.
The \textbf{frontend} runs on the DPU's ARM cores: it receives client requests~\comp{1}, tokenizes prompts~\comp{2}, locates a free ring buffer slot~\comp{3}\comp{4}, writes prompts into GPU memory via one-sided RDMA~\comp{5}, retrieves generated tokens~\comp{11}, detokenizes them~\comp{12}, and streams responses back to clients~\comp{13}.
The \textbf{backend} runs entirely on the GPU: a persistent scheduler kernel polls a shared ring buffer for incoming prompts~\comp{6}, selects and launches pre-captured CUDA graphs~\comp{7}\comp{8}, computes tokens from the logits via on-device sampling~\comp{9}, and publishes results back to the ring buffer~\comp{10}---all without returning control to the host.
The two planes communicate exclusively through a \emph{GPU-resident ring buffer} accessed by RDMA (\comp{5},~\comp{11}), which serves as the sole rendezvous point and eliminates any need for host-mediated coordination.
This cleanly separates \emph{initialization} (host-assisted) from \emph{steady state} (DPU+GPU only).

\smartparagraph{Why a DPU--GPU split?}
Modern SmartNICs outperform server-class CPUs (Xeons, EPYCs) in network bandwidth per core by $15{-}24\times$ and memory bandwidth per core by $1.6{-}2.1\times$~\cite{lovelock}, making them well suited to transport and request management tasks.
However, their limited core count cannot absorb the volume of per-iteration scheduling decisions in continuous batching without becoming a bottleneck.
\systemname exploits this asymmetry: the DPU handles request management and network I/O, while GPU-resident persistent kernels handle the latency-critical scheduling and execution loop.
This split removes the host CPU from the data path without imposing DPU-side bottlenecks.

\subsection{Backend: GPU-Resident Inference Engine}
\label{sec:design-backend}

The GPU integrates four subsystems: \textbf{(1) a persistent GPU scheduler} that continuously manages request lifecycles, \textbf{(2) device-side CUDA graph launch} that avoids host-mediated kernel dispatch, \textbf{(3) a continuous batching engine} that admits new requests into running decode batches, and \textbf{(4) a lock-free ring buffer} for DPU--GPU coordination.

\smartparagraph{Persistent scheduler.}
\label{sec:design-persistent-sched}
\systemname replaces the host-driven decode loop (\S\ref{sec:host-bottleneck}) with a single persistent CUDA kernel that occupies one thread block (256 threads) and runs indefinitely.
The scheduler executes an infinite control loop: (1) it scans the ring buffer for newly submitted prompts, (2) claims them via atomic compare-and-swap, (3) selects and launches the appropriate CUDA graph for prefill or decode, (4) polls device-resident output buffers for completion after token sampling, and (5) publishes generated tokens and status updates back to the ring buffer.
Since the scheduler never yields to the host, there is no kernel launch overhead, no host--device synchronization, and no CPU-mediated bookkeeping on the token-critical path.

\smartparagraph{Parallel slot scanning.}
To avoid serialization, all 256 threads in the scheduler's thread block scan disjoint contiguous ranges of the ring buffer's slots in parallel and claim pending slots via atomic CAS, eliminating locks and inter-block synchronization.
A complete scan of all 4096 slots completes in \qtyrange{1}{5}{\microsecond}, ensuring that prompt admission latency is dominated by RDMA transfer time rather than scheduling overhead.

\begin{figure}[t]
  \centering
  \includegraphics[width=\columnwidth]{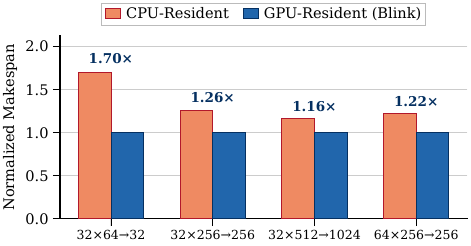}
  \caption{Normalized makespan: GPU-resident vs.\ CPU-resident scheduling on Qwen3-32B (N$\times$I$\rightarrow$O denotes N requests, I input tokens, O output tokens).}
  \label{fig:cpu-vs-gpu-scheduler}
  \Description[CPU vs.\ GPU-resident scheduler makespan]{Bar chart comparing normalized makespan. The CPU-resident scheduler is 1.16--1.70x slower across four workload configurations.}
\end{figure}

Fig.~\ref{fig:cpu-vs-gpu-scheduler} quantifies the benefit of GPU-resident scheduling. We run identical workloads on the same compiled engine (Qwen3-32B, FP16, H100, batch size 16) under two scheduler placements that share an identical scheduling policy. In the CPU-resident baseline, the sampled tokens are copied to host memory over PCIe after each decode step, and the batch is reassembled on the CPU before launching the next graph; in contrast, the GPU-resident scheduler eliminates this per-step synchronization. In both cases, token sampling is performed on the GPU to best match popular CPU-centric systems such as vLLM. Across four workload configurations, the CPU path inflates the total makespan by $1.16{-}1.70\times$, with the largest penalty on short-output workloads where per-step PCIe round-trip overhead dominates compute.

\smartparagraph{Device-side CUDA graph launch.}
\label{sec:design-graph-launch}
A key enabler of CPU-free operation is \emph{device-side graph launch}~\cite{cuda_device_graph_launch, nvidia_dgl_blog}: the persistent scheduler, itself wrapped as a CUDA graph, enqueues pre-captured inference graphs for execution directly from the device.
\systemname uses the fire-and-forget launch mode, which completes in ${\approx}$\qty{2}{\microsecond} across representative graph topologies (straight-line, fork-join, and parallel)---\numrange{5}{8}$\times$ faster than host-side launch (\qtyrange{11}{17}{\microsecond} depending on topology) and \num{2.7}$\times$ faster than tail launch (${\approx}$\qty{5.5}{\microsecond}).
Over hundreds of decode steps these differences compound: fire-and-forget saves \qtyrange{4.6}{7.7}{\milli\second} per 512-token generation compared to host launch.
More critically, host launch reintroduces the CPU onto the critical path, re-exposing inference to host-side interference.

\smartparagraph{Challenge: the 120-launch hard limit.}
Fire-and-forget imposes a fundamental constraint that is not widely documented: the CUDA runtime limits outstanding fire-and-forget launches from a single parent graph execution to 120~\cite{cuda_fire_and_forget}; once this limit is reached, further launches produce undefined behavior.
For an inference server that must generate tokens indefinitely, this is a critical obstacle---a single request with 512 output tokens would exhaust the launch budget well before completion.
Falling back to host launch after 120 iterations would reintroduce the CPU onto the critical path, and using tail launch exclusively incurs $2.7\times$ higher per-launch overhead that accumulates across hundreds of decode steps.

\smartparagraph{Solution: window-based tail-launch recovery.}
\systemname resolves this constraint with a \emph{window-based tail-launch recovery} mechanism.
The scheduler maintains a monotonically increasing launch counter in shared memory.
Upon reaching the 120-launch limit, it issues a single tail launch that atomically replaces the current graph execution with a fresh instance.
All state---ring buffer pointers, KV cache metadata, in-flight requests---resides in persistent GPU memory and survives graph re-instantiation; the new instance resets the counter to zero and resumes the scheduling loop from the same logical point.
This design achieves three properties:
\first~\emph{near-optimal launch latency}---fire-and-forget is used for 120 of every 121 iterations, with a single tail launch amortized across the window ({<}\qty{0.03}{\microsecond} overhead per decode step);
\Second~\emph{seamless state continuity}---all metadata, KV cache state, and ring buffer contents reside in persistent GPU memory that survives graph re-instantiation; and 
\third~\emph{unbounded generation}---no upper bound on the number of generated tokens.

\smartparagraph{Completion detection.}
Fire-and-forget launches provide no host-side completion callback---the parent kernel receives no notification when a child graph finishes.
\systemname performs \emph{polling-based completion detection} entirely on the device: after the inference engine executes a forward pass and token sampling writes the result, the scheduler polls the designated output buffer until the extracted token appears.
For prefill, the scheduler polls a token extraction buffer populated by the inference graph upon computing the first output token; buffer occupancy signals prefill completion.
For decode, the scheduler polls per-step extraction buffers written by each decode graph.
The polling cost is negligible (a few shared-memory loads per iteration) relative to inference compute, and it preserves the invariant that no host involvement occurs during steady state.

\smartparagraph{Continuous batching with inline prefill.}
\label{sec:design-batching}
\systemname implements continuous batching~\cite{orca} with FCFS scheduling, matching the policy used by TensorRT-LLM, vLLM~\cite{vllm}, and SGLang~\cite{sglang}, so that evaluation differences (\S\ref{sec:evaluation}) isolate the architectural benefit of removing the host CPU rather than a scheduling advantage.
To admit new requests without stalling ongoing generation, the scheduler implements \emph{pause-and-resume batching} with an overlapped ring scan: while a decode graph executes asynchronously, the scheduler's 256 threads scan the ring buffer in parallel for pending prompts, \emph{hiding scan latency behind decode compute}.
If candidates are found, the scheduler evaluates three conditions before pausing: \first~pending prefills detected during the overlapped scan, \Second~free batch-slot capacity (accounting for slots that will complete this step), and \third~sufficient fire-and-forget launch-window headroom for the prefill graph plus resumed decode.
When all three hold, the scheduler pauses in-flight decode requests after the current step, executes a prefill graph for the new arrivals, merges the newly prefilled requests into the decode batch, and restarts the decode loop---all within a single persistent kernel invocation, with no host round-trip.
This ensures that new requests are admitted within one decode step's latency, bounding \gls{TTFT} without sacrificing decode throughput.

\smartparagraph{Ring buffer.}
The ring buffer resides in GPU memory and is the only shared data structure between the DPU and GPU, serving as the target of one-sided RDMA reads\slash writes.
It consists of a fixed set of slots plus shared arenas for input and generated tokens.
Each slot records per-request metadata (\eg prompt identity, token counts, and generation progress) and offsets into the token arenas.
The scheduler advances each slot through a lifecycle state machine---\textsc{empty} $\to$ \textsc{prefill\_pending} $\to$ \textsc{prefill\_processing} $\to$ \textsc{decode\_processing} $\to$ \textsc{decode\_completed} $\to$ \textsc{empty}---and uses a \textsc{decode\_paused} state to support preemption and continuous batching.
Ownership and state transitions use atomic Compare-and-Swap (CAS).
The layout and update rules are designed to avoid intermediate copying, tolerate benign races, and ensure that RDMA-visible updates become visible in the intended order via memory fences.

\smartparagraph{CUDA graph cache.}
\label{sec:design-engine}
\systemname uses precompiled TensorRT~\cite{tensorrt} engines for inference.
During initialization, the host captures inference computation as CUDA graphs for a dense grid of (batch size, sequence length) pairs and instantiates each for device-side launch so the persistent scheduler can invoke them directly.
All graphs reuse a single set of device buffers allocated once for the maximum supported shape, avoiding per-graph memory duplication.
Since each captured graph consumes only \qtyrange{2}{3}{MB}---orders of magnitude smaller than the hundreds of MBs typical of PyTorch-based CUDA graphs---a cache of $650$–$1000$ graphs fits within \qtyrange{2}{4}{GB} in our evaluated models, enabling fine-grained shape matching without exhausting the memory budget and overhead of dynamic capturing.

At runtime, the scheduler selects the tightest-fitting prefill graph via a precomputed lookup table indexed by (batch, sequence length), achieving $O(1)$ selection with no per-step search; a maximum-shape fallback graph handles any combination not in the cache.
Token sampling (Top-P with temperature) is captured \emph{inside} each graph, so the entire forward pass from attention through next-token selection executes as a single device-side launch with no host round-trip.

\subsection{Model Compatibility}
\label{sec:design-compatibility}

Because the persistent scheduler treats the inference graph as an opaque computation---populating input tensors, launching the graph, and reading output buffers---\systemname inherits TensorRT's broad model support without modification.
Dense Transformers, \gls{MoE} models, and encoder-decoder architectures require only a TensorRT compilation step; no changes to the scheduler, ring buffer, or RDMA datapath are needed.
Both model-level innovations (\eg new attention mechanisms~\cite{mqa,gqa}, quantization schemes, activation functions) and serving-level optimizations (chunked prefill~\cite{sarathi_serve}, prefix caching~\cite{sglang}, speculative decoding~\cite{spec-decoding}) are orthogonal to \systemname's architecture; we discuss concrete integration paths in \S\ref{sec:discussion}.

\subsection{Frontend: DPU-Side Control and Data Plane}
\label{sec:design-frontend}

The frontend runs on the BlueField DPU's ARM cores and manages the full request lifecycle from arrival through token delivery.
A \emph{request tracker} maintains per-request state---slot assignment, token counts, and completion status---coordinating prompt submission, token retrieval, and client-facing streaming across the subsystems described below.
The frontend also hosts a thin OpenAI-compatible HTTP server with \gls{SSE} streaming support; owing to the decoupled architecture, this interface layer could alternatively run on a separate gateway.

\smartparagraph{RDMA datapath.}
The frontend uses one-sided RDMA to directly read and write the GPU-resident ring buffer.
It stages outgoing prompts and incoming results in separate DPU-local buffers to decouple submission from retrieval and to reduce interference between control metadata and bulk token traffic.
On submission, the frontend transfers the tokenized prompt into the assigned slot and updates its state so the backend can begin prefill.
When requests arrive in bursts, the frontend coalesces transfers to amortize RDMA overhead across multiple prompts.

\smartparagraph{Slot tracker.}
Rather than scanning all ring buffer slots via RDMA before each submission, the slot tracker maintains a local availability cache on the DPU, refreshed periodically via a single bulk RDMA read.
A hint-based circular scan finds empty slots in $O(1)$ amortized time, improving spatial locality and reducing per-submission RDMA overhead.

\smartparagraph{Token reader.}
A background token reader continuously polls the ring buffer for generated tokens.
Each cycle, it issues one RDMA read to refresh cached slot metadata (64\,KB), then compares each active slot's generation count with its local state to detect new output.
To minimize TTFT, new slots go to an \emph{urgent slot} scanned first, so the first token is retrieved within one poll interval.
Adaptive polling bounds per-token latency while limiting RDMA traffic.
Under bursty arrivals, the reader caps per-poll work and uses large RDMA task pools to avoid completion-queue saturation; a dedicated progress thread processes completions to sustain throughput.
Retrieved tokens go to the detokenizer and are streamed to clients via SSE.

\begin{figure}[t]
  \centering
  \includegraphics[width=\columnwidth]{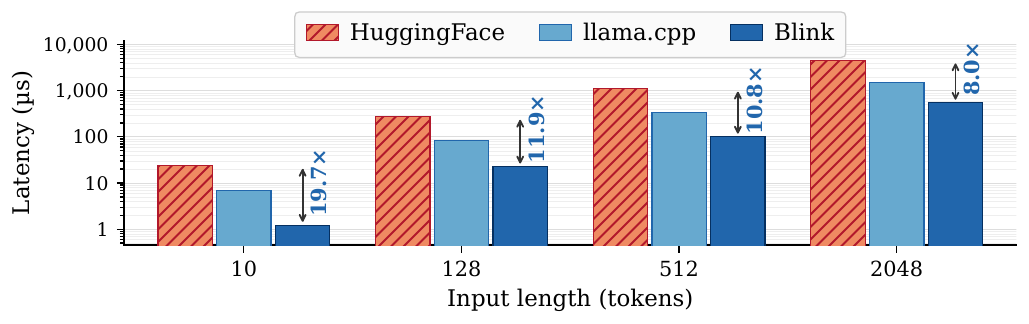}
  \caption{Tokenization latency on BlueField-3 ARM A78 cores (\systemname) vs.\ Intel Xeon (HuggingFace, llama.cpp). Labels show speedup relative to HuggingFace.}
  \label{fig:tokenizer-latency}
  \Description[Tokenization latency comparison]{Bar chart comparing tokenization latency for HuggingFace, llama.cpp, and \systemname across input lengths of 10--2048 tokens. \systemname is 8--19.7x faster than HuggingFace.}
\end{figure}

\smartparagraph{Tokenizer.}
Tokenization on the DPU must be efficient on the ARM microarchitecture.
\systemname implements a tokenizer with merge rules in a 64-byte-aligned flat hash table, packing four key-value pairs per L1D cache line; 16-byte-aligned symbol nodes keep short-word working sets within two cache lines, avoiding heap indirection.
Regex pre-tokenization uses ARM NEON SIMD for byte classification at 16~bytes per cycle, and all per-request state lives in pre-allocated thread-local buffers, eliminating heap allocation on the request path.
Fig.~\ref{fig:tokenizer-latency} compares \systemname's tokenizer on BlueField-3 ARM A78 cores with HuggingFace's (used by vLLM and SGLang) and llama.cpp's~\cite{llamacpp} tokenizers on an Intel Xeon CPU.
Despite the ARM cores’ lower clock speed, \systemname's tokenizer is $8{-}19.7\times$ faster than HuggingFace for inputs of \numrange{10}{2048} tokens and consistently outperforms llama.cpp, showing that the DPU introduces no tokenization bottleneck.

\section{Implementation}
\label{sec:implementation}
\systemname's GPU-resident backend comprises approximately \num{16}\,k lines of CUDA and C++ code implementing the persistent scheduler, device-side graph launch mechanism, continuous batching engine, KV-cache management, and token sampling, targeting CUDA 13.1 and leveraging the TensorRT inference engine for pre-captured model graphs, while the DPU-resident control and data plane adds approximately \num{17}\,k lines of C/C++ code implementing HTTP parsing and request validation, RDMA orchestration via the DOCA 3.2 SDK, ring buffer coordination, and \gls{SSE} streaming, with the DPU also providing optional tokenization on its ARM cores to offload this traditionally host-side work and reduce end-to-end latency.


\section{Evaluation}
\label{sec:evaluation}
We evaluate \systemname against three state-of-the-art \gls{LLM} inference serving systems across four models, under both isolated and co-located multi-tenant conditions.
We address three questions:

\begin{enumerate}[leftmargin=*,nosep,label=Q\arabic*]
    \item How does \systemname compare to existing inference systems in latency and throughput? (\S\ref{sec:eval-normal})
    \item How well does \systemname preserve performance under co-location with noisy tenants? (\S\ref{sec:eval-interference})
    \item How does \systemname's architecture affect energy efficiency under isolated and multi-tenant execution? (\S\ref{sec:eval-energy})
\end{enumerate}
\noindent We report P99 \gls{TTFT}, P99 \gls{TPOT}, and throughput (completed req/s), emphasizing tail latencies as they determine whether \glspl{SLO} can be met in production.
Across all three questions, the pattern is consistent: in isolation,
\systemname provides the best pre-saturation latency envelope and the highest
saturated throughput; under CPU interference, its operating range and plateau
remain essentially unchanged while CPU-mediated baselines collapse; and
because all systems draw similar wall power, these throughput gains translate
directly into lower energy per token.

\subsection{Experimental Setup}
\label{sec:eval-setup}
\smartparagraph{Testbed.}
Table~\ref{tab:system-config} lists the hardware. The \systemname backend runs on this server; its frontend runs on a separate BlueField-3 DPU machine, connected via a 200\,Gbps RDMA link (DOCA SDK v3.2.1). A workload generator reaches the frontend through a 10\,Gbps switch. Baselines run on the same server with the workload generator connected via the same switch.

\begin{table}[ht]
\centering
\caption{Hardware configuration.}
\label{tab:system-config}
\resizebox{.85\linewidth}{!}{%
\begin{tabular}{ll}
\textbf{Component} & \textbf{Specification} \\
\midrule
GPU & NVIDIA H100 (96\,GB HBM3) \\
CPU & 2 $\times$ Intel Xeon Gold 6336Y (96 cores @ 2.40\,GHz),\\
& DVFS disabled, governor: \texttt{performance} \\
DRAM & 256\,GB DDR5 \\
Network & ConnectX-6 (200\,Gbps) \\
DPU & BlueField-3 (16 ARM Cortex-A78, 32\,GB) \\
OS & Linux 5.15 (Ubuntu 22.04 LTS) \\
\end{tabular}
}
\end{table}

\smartparagraph{Models and workloads.}
We evaluate four models that span a range of parameter counts and architectural choices:
Llama-3 8B~\cite{llama3} (dense, 8B parameters),
Phi-4 15B~\cite{phi4} (dense, 14.7B parameters),
Qwen-3 32B~\cite{qwen3} (dense, 32B parameters), and
Qwen-3 30B-A3B~\cite{qwen3} (\acrshort{MoE}, 30B total\slash 3B active parameters).
All experiments use paged attention and fp16 precision.
We use the \textit{guidellm} tool~\cite{guidellm} configured with the ShareGPT v3 dataset~\cite{sharegpt_v3}, which provides naturalistic conversation traces.
For each experiment we sweep over 13 offered load levels from 1 to 32 requests/second, measuring performance at each rate for \qty{60}{\second} before advancing to the next.
Both \systemname and TensorRT-LLM use fp16 precision with paged KV cache, float16 attention (context FMHA), fused MLP, and removed input padding; the MoE plugin is enabled (fp16) only for Qwen-3 30B\nobreakdash-A3B.

\smartparagraph{Baselines.}
We compare against three production-grade systems:
(1)~\textit{TensorRT-LLM v1.1.0}~\cite{tensorrt_llm} with TensorRT v10.14~\cite{tensorrt} as its execution backend---using the \emph{same} TensorRT engines as \systemname to isolate the effect of CPU-mediated orchestration;
(2)~\textit{vLLM v0.13.0}~\cite{vllm}, the widely deployed open-source serving system; and
(3)~\textit{SGLang v0.5.8}~\cite{sglang}, a recent high-performance serving system with RadixAttention.
All baselines use recommended production settings with CUDA Graphs; chunked prefill, prefix caching, and CPU offloading are disabled for controlled comparison, as \systemname does not yet incorporate these optimizations.

\smartparagraph{Interference workloads.}
To model multi-tenant ``noisy neighbor'' scenarios, we colocate two CPU-intensive workloads with inference: \textit{pbzip2}~\cite{pbzip2} compressing a 50\,GB file with 45 threads, and \textit{Ninja}~\cite{ninja} building the LLVM~\cite{llvm} compiler source with 45 parallel jobs. Following NVIDIA's guidelines~\cite{nvidia_cert_guide}, we reserve six cores for inference and leave the remaining 90 host cores to the interferers. We also tested ffmpeg video encoding and observed qualitatively similar results, so we omit those experiments for brevity.

\subsection{Isolated Performance (Q1)}
\label{sec:eval-normal}

We first evaluate under isolated execution which is the best-case scenario for
CPU-mediated systems, where inference has exclusive access to host resources.
Rather than centering the evaluation on a single unloaded rate, we summarize
performance over the \emph{Non saturated operational range}: for each model, the
set of offered loads up to \systemname's throughput saturation point, where the
throughput curve transitions from linear growth to a stable plateau.  This
asks the deployment question that matters: over the full load range that
\systemname can still absorb before entering saturation, how do competing
systems behave?  This framing is intentional rather than favorable: if a
baseline already queues inside the load interval that \systemname can still
sustain, that is precisely the deployment disadvantage the paper aims to
measure.

We identify the saturation point from the same rate-averaged throughput curves
in Fig.~\ref{fig:throughput}(a--d) using a two-segment fit (linear growth
followed by plateau).  To match the figures, we first average repeated runs
at each offered load and then aggregate across loads.
The resulting \systemname operating ranges, where \(\lambda\) denotes the offered request rate in req/s, are \(\lambda \le 12\) for
Llama-3 8B, \(\lambda \le 7\) for Phi-4 15B, \(\lambda \le 2\) for Qwen-3 32B,
and \(\lambda \le 4\) for Qwen-3 30B-A3B.  Table~\ref{tab:isolated-summary}
reports geometric-mean P99 TTFT and P99 TPOT over the corresponding
rate-averaged latency curves in Figs.~\ref{fig:p99-latency}(a--d)
and~\ref{fig:p99-latency}(e--h), together with throughput at the offered load
equal to \systemname's saturation point from Fig.~\ref{fig:throughput}(a--d).

\begin{table}[t]
\centering
\setlength{\tabcolsep}{2pt}
\caption{Pre-saturation summary over the \systemname-defined operating range.
Repeated runs are averaged per load to match the figures. TTFT and TPOT are
in milliseconds; Tput@sat is achieved throughput at \systemname's saturation
point.  Best values per model 
are boldfaced.}
\label{tab:isolated-summary}
\resizebox{0.65\linewidth}{!}{%
\begin{tabular}{llrrr}
\textbf{Model} & \textbf{System} & \shortstack[c]{\textbf{Geo.\ P99}\\\textbf{TTFT}} & \shortstack[c]{\textbf{Geo.\ P99}\\\textbf{TPOT}} & \shortstack[c]{\textbf{Tput}\\\textbf{at sat.}} \\
\midrule
\multirow{4}{*}{\shortstack[l]{Llama-3\\8B\\$\lambda \le 12$}}
  & \systemname &   \textbf{653.8} &  \textbf{15.1} & \textbf{11.87} \\
  & TRT-LLM     &   880.0 &  17.7 & 10.80 \\
  & vLLM        &  1309.6 &  24.2 &  9.12 \\
  & SGLang      &  1747.1 &  30.7 &  7.88 \\
\midrule
\multirow{4}{*}{\shortstack[l]{Phi-4\\15B\\$\lambda \le 7$}}
  & \systemname &  \textbf{1109.4} &  \textbf{25.0} &  \textbf{6.72} \\
  & TRT-LLM     &  1453.8 &  29.8 &  6.42 \\
  & vLLM        &  1683.7 &  34.5 &  6.05 \\
  & SGLang      &  2874.1 &  47.9 &  5.58 \\
\midrule
\multirow{4}{*}{\shortstack[l]{Qwen-3\\32B\\$\lambda \le 2$}}
  & \systemname &  \textbf{9481.3} & \textbf{113.4} & \textbf{2.00} \\
  & TRT-LLM     &  9621.4 & 115.2 & 1.97 \\
  & vLLM        & 10862.4 & 133.7 & 1.88 \\
  & SGLang      & 11413.0 & 123.3 & 1.85 \\
\midrule
\multirow{4}{*}{\shortstack[l]{Qwen-3\\30B-A3B\\$\lambda \le 4$}}
  & \systemname &  \textbf{1397.5} &  \textbf{35.5} & \textbf{4.85} \\
  & TRT-LLM     &  4814.7 &  65.8 & 3.61 \\
  & vLLM        &  8919.2 &  90.9 & 2.91 \\
  & SGLang      & 11839.8 & 120.8 & 2.62 \\
\end{tabular}
}
\end{table}

\smartparagraph{Within \systemname's serviceable range, \systemname provides the best latency envelope.}
Table~\ref{tab:isolated-summary} shows that across the entire load range \systemname
can sustain before saturation, \systemname delivers the lowest
geometric-mean P99 TTFT and TPOT on three of the four models and is
near-parity with TensorRT-LLM on the strongly GPU-bound Qwen-3 32B.  On
Llama-3 8B, baselines incur \numrange{1.35}{2.67}$\times$ higher
geometric-mean P99 TTFT and \numrange{1.17}{2.03}$\times$ higher
geometric-mean P99 TPOT.  On Phi-4 15B, the corresponding gaps are
\numrange{1.31}{2.59}$\times$ for TTFT and \numrange{1.19}{1.92}$\times$ for
TPOT.  The \acrshort{MoE} result is especially strong: on Qwen-3 30B-A3B,
baselines incur \numrange{3.45}{8.47}$\times$ higher TTFT and
\numrange{1.85}{3.40}$\times$ higher TPOT than \systemname.  Throughput at saturation
makes the serviceability difference explicit: \systemname sustains
\qty{11.87}{req/s} versus \qty{10.80}{req/s} for TensorRT-LLM on Llama-3 8B,
\qty{6.72}{req/s} versus \qty{6.42}{req/s} on Phi-4 15B, and
\qty{4.85}{req/s} versus \qty{3.61}{req/s} on the \acrshort{MoE} model.

Two architectural properties account for the TTFT advantage. First, the persistent GPU scheduler continuously scans the ring buffer with 256 threads in parallel and can claim a newly submitted prompt within \qtyrange{1}{5}{\microsecond} (\S\ref{sec:design-persistent-sched}), whereas CPU-mediated systems incur host-side scheduling and kernel dispatch delays of tens to hundreds of microseconds per step. Second, prompt tokens reach GPU memory via zero-copy one-sided RDMA from the DPU, bypassing the host memory hierarchy entirely; in contrast, all baselines copy input tokens through host DRAM before transferring them to the GPU. Together, these properties reduce the time between prompt arrival and the start of prefill computation, which is the dominant contributor to TTFT at pre-saturation loads.

\begin{figure}[t]
    \centering
    \raisebox{12mm}{\rotatebox{90}{\small Latency (ms)}}%
    \begin{subfigure}[b]{0.46\columnwidth}
        \centering
        \includegraphics[width=\linewidth]{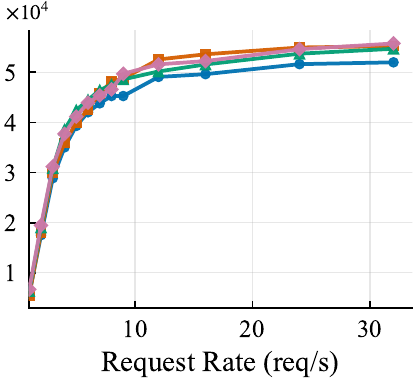}
        \caption{P99.9 TTFT}
    \end{subfigure}\hfill
    \begin{subfigure}[b]{0.46\columnwidth}
        \centering
        \includegraphics[width=\linewidth]{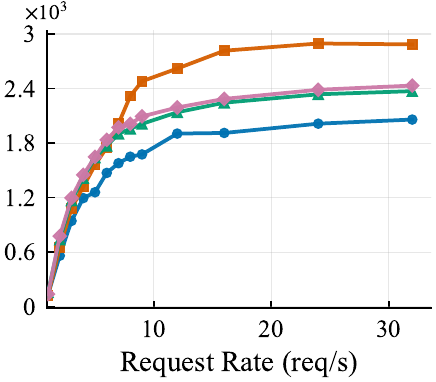}
        \caption{P99.9 TPOT}
    \end{subfigure}
    \figlegendnosuffix
    \caption{P99.9 tail latency for Qwen-3 32B (isolated).
    Although P99 latencies are near-parity, P99.9 reveals a consistent
    \systemname advantage across all baselines that grows with load.}
    \label{fig:p999-qwen32b}
    \Description[P99.9 TTFT and TPOT for Qwen-3 32B]{P99.9 TTFT and TPOT latency comparison between \systemname, TensorRT-LLM, vLLM, and SGLang on Qwen-3 32B under isolated execution.}
\end{figure}

\begin{figure*}[t]
    \centering
    \raisebox{16mm}{\rotatebox{90}{\small P99 TTFT (ms)}}%
    \begin{subfigure}[b]{0.240\textwidth}
        \centering
        \includegraphics[width=\linewidth]{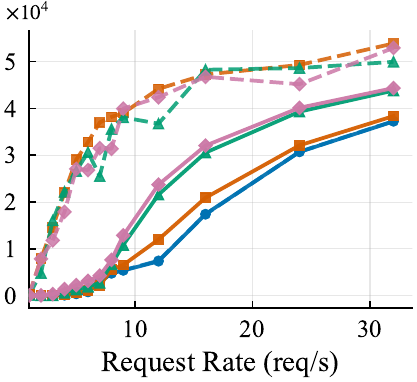}
        \caption{Llama-3 8B}
    \end{subfigure}\hfill
    \begin{subfigure}[b]{0.240\textwidth}
        \centering
        \includegraphics[width=\linewidth]{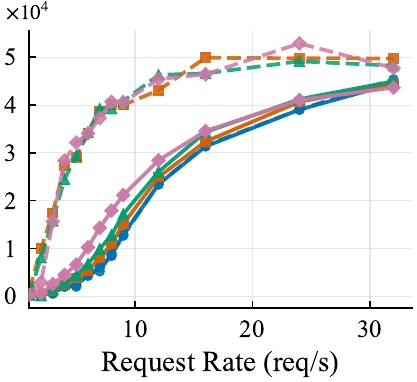}
        \caption{Phi-4 15B}
    \end{subfigure}\hfill
    \begin{subfigure}[b]{0.240\textwidth}
        \centering
        \includegraphics[width=\linewidth]{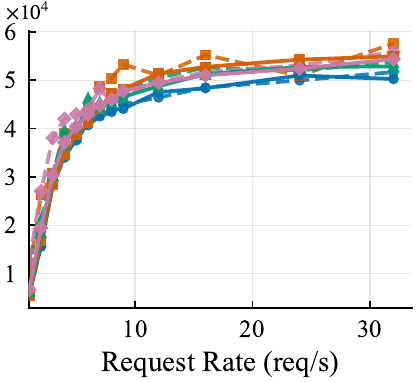}
        \caption{Qwen-3 32B}
    \end{subfigure}\hfill
    \begin{subfigure}[b]{0.240\textwidth}
        \centering
        \includegraphics[width=\linewidth]{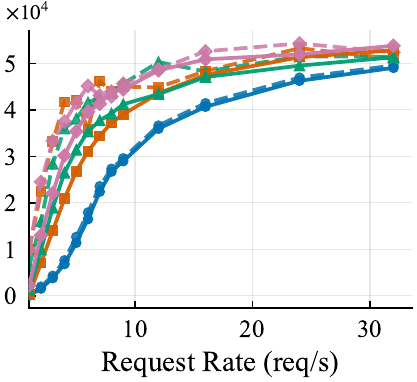}
        \caption{Qwen-3 30B-A3B}
    \end{subfigure}\\[4pt]
    \raisebox{16mm}{\rotatebox{90}{\small P99 TPOT (ms)}}%
    \begin{subfigure}[b]{0.240\textwidth}
        \centering
        \includegraphics[width=\linewidth]{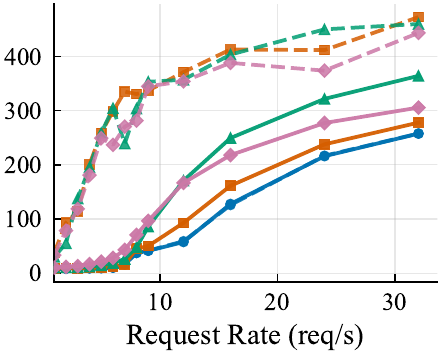}
        \caption{Llama-3 8B}
    \end{subfigure}\hfill
    \begin{subfigure}[b]{0.240\textwidth}
        \centering
        \includegraphics[width=\linewidth]{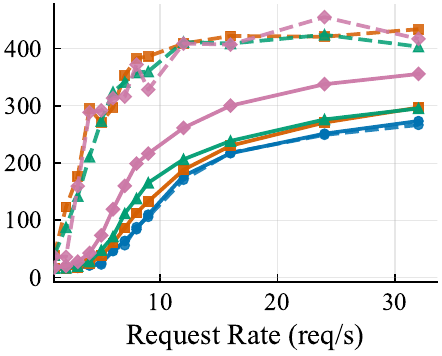}
        \caption{Phi-4 15B}
    \end{subfigure}\hfill
    \begin{subfigure}[b]{0.240\textwidth}
        \centering
        \includegraphics[width=\linewidth]{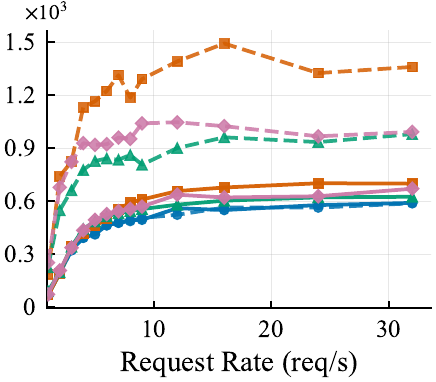}
        \caption{Qwen-3 32B}
    \end{subfigure}\hfill
    \begin{subfigure}[b]{0.240\textwidth}
        \centering
        \includegraphics[width=\linewidth]{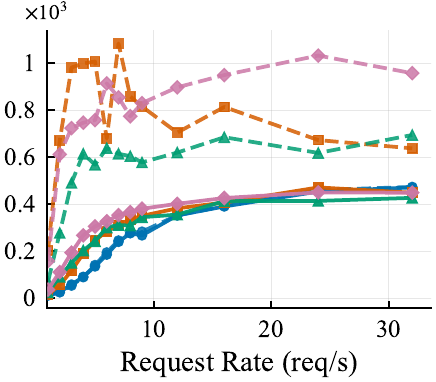}
        \caption{Qwen-3 30B-A3B}
    \end{subfigure}
    \figlegend
    \caption{P99 tail latency across four models. Top row~(a--d): TTFT; bottom row~(e--h): TPOT. Solid lines show isolated execution; dashed lines show CPU interference. Within \systemname's operating range, \systemname maintains a flatter latency envelope and near-overlapping isolated and interference curves, whereas baseline systems queue earlier and suffer severe tail inflation under co-location.}
    \label{fig:p99-latency}
    \Description[P99 TTFT and TPOT under isolated and interference conditions]{P99 TTFT and TPOT latency across model architectures under both isolated and interference conditions.}
\end{figure*}

\smartparagraph{The same-engine comparison isolates the orchestration benefit.}
The comparison to TensorRT-LLM is especially clean because both systems use
the same TensorRT engines.  Across \systemname's operating range, \systemname
reduces geometric-mean TTFT by \num{1.35}$\times$ on Llama-3 8B,
\num{1.31}$\times$ on Phi-4 15B, and \num{3.45}$\times$ on Qwen-3 30B-A3B.
These gains therefore come from removing host-mediated orchestration rather
than changing the model backend.  Qwen-3 32B is the expected limiting case:
its operating range is narrow and the P99 gap compresses, confirming that once
the workload becomes predominantly GPU-bound there is less host-side latency
left to remove.  However, the advantage re-emerges at deeper tail percentiles:
Fig.~\ref{fig:p999-qwen32b} plots P99.9 TTFT and TPOT for all four systems
on Qwen-3 32B.  At P99.9, baselines incur \qtyrange{4}{8}{\percent} higher
TTFT and \qtyrange{15}{48}{\percent} higher TPOT than \systemname across
saturated loads---a clear separation that P99 averages mask.  Supplementary
materials contain corresponding P99.9 breakdowns for all four models.

\smartparagraph{Beyond saturation, \systemname sustains the highest plateau.}
The throughput curves in Fig.~\ref{fig:throughput}(a--d) show that
\systemname also reaches the latest or tied-latest saturation point and then
sustains the highest plateau throughput on every model.  The plateau is \qty{11.96}{req/s}
for Llama-3 8B, \qty{6.86}{req/s} for Phi-4 15B, \qty{2.13}{req/s} for
Qwen-3 32B, and \qty{5.07}{req/s} for Qwen-3 30B-A3B.  Relative to
TensorRT-LLM, this corresponds to \qty{9}{\percent}, \qty{6}{\percent},
\qty{8}{\percent}, and \qty{37}{\percent} higher plateau throughput,
respectively.  The \acrshort{MoE} model consistently exhibits the largest gain
across all metrics; two reinforcing factors explain why.

\smartparagraph{Why the \acrshort{MoE} advantage is amplified.}
First, \acrshort{MoE} models have an unfavorable \emph{compute-to-orchestration
ratio}: Qwen-3 30B-A3B activates only \qty{3}{B} of its \qty{30}{B} parameters
per token, so each decode step completes quickly on the GPU---comparable in
compute to a small dense model---yet the per-step CPU orchestration cost
(scheduler iteration, host--device synchronization, batch reassembly) remains
constant.  CPU overhead therefore consumes a much larger fraction of total step
time than in dense models where GPU compute dominates, and removing that
overhead yields a proportionally larger speedup.  This explains the progression
from \qty{8}{\percent} on Qwen-3 32B (GPU-bound; CPU overhead is a small
fraction) to \qty{37}{\percent} on Qwen-3 30B-A3B (fast active compute; CPU
overhead is a large fraction).
Second, \acrshort{MoE} expert routing is \emph{data-dependent but not
shape-dependent}: which experts are selected varies with each token's hidden
state, but all tensor dimensions remain fixed regardless of the routing
decision.  TensorRT's \acrshort{MoE} plugin handles gating, token-to-expert
dispatch, and gather internally within fixed-size buffers, so the entire
forward pass---including \acrshort{MoE} layers---is captured as a single CUDA
graph with static shapes.  \systemname's device-side graph launch therefore
executes \acrshort{MoE} models without any host intervention to interpret
router outputs or dynamically dispatch expert kernels.  CPU-mediated baselines,
by contrast, still interpose host-side scheduling on every decode step, paying
the same per-step orchestration tax even though the compiled engine itself
could run autonomously.

\vspace{-1em}
\subsection{Performance Under CPU Interference (Q2)}
\label{sec:eval-interference}

Production deployments routinely co-locate inference with other tenants.  We
evaluate whether \systemname's host-decoupled execution preserves the same
operating range under severe CPU contention.  We deliberately keep the ranges
fixed to the isolated \systemname saturation points from \S\ref{sec:eval-normal}: this tests
whether the load range that \systemname unlocks in isolation remains usable under
colocation, and whether competing systems can match that behavior.  If a
baseline cannot sustain that same range once interference appears, that loss
of serviceable capacity is itself the result.

\begin{table}[t]
\centering
\setlength{\tabcolsep}{1pt}
\caption{Same pre-saturation summary as Table~\ref{tab:isolated-summary}, but
under CPU interference.  Bracketed terms show interference\,/\,isolation;
throughput brackets report retention at \systemname's saturation point.  Raw
latencies are in milliseconds and throughput in requests/s.  Best raw values per model 
are boldfaced.}
\label{tab:interference-summary}
\resizebox{0.8\linewidth}{!}{%
\begin{tabular}{llccc}
\textbf{Model} & \textbf{System} & \shortstack[c]{\textbf{Geo.\ P99}\\\textbf{TTFT}} & \shortstack[c]{\textbf{Geo.\ P99}\\\textbf{TPOT}} & \shortstack[c]{\textbf{Tput}\\\textbf{at sat.}} \\
\midrule
\multirow{4}{*}{\shortstack[l]{Llama-3\\8B\\$\lambda \le 12$}}
  & \systemname & \textbf{652.8 [1.00]} & \textbf{15.2 [1.00]} & \textbf{11.83 [1.00]} \\
  & TRT-LLM     & 16574 [18.84] & 196.6 [11.10] & 4.13 [0.38] \\
  & vLLM        & 14563 [11.12] & 178.2 [7.35] & 4.00 [0.44] \\
  & SGLang      & 14728 [8.43] & 177.3 [5.77] & 3.78 [0.48] \\
\midrule
\multirow{4}{*}{\shortstack[l]{Phi-4\\15B\\$\lambda \le 7$}}
  & \systemname & \textbf{1023.2 [0.92]} & \textbf{24.4 [0.98]} & \textbf{6.77 [1.01]} \\
  & TRT-LLM     & 15496 [10.66] & 183.7 [6.17] & 2.62 [0.41] \\
  & vLLM        & 12016 [7.14] & 163.5 [4.74] & 2.87 [0.47] \\
  & SGLang      & 10991 [3.82] & 150.6 [3.15] & 2.62 [0.47] \\
\midrule
\multirow{4}{*}{\shortstack[l]{Qwen-3\\32B\\$\lambda \le 2$}}
  & \systemname & \textbf{9415.6 [0.99]} & \textbf{117.8 [1.04]} & \textbf{2.05 [1.02]} \\
  & TRT-LLM     & 16195 [1.68] & 371.4 [3.23] & 1.00 [0.51] \\
  & vLLM        & 16702 [1.54] & 353.0 [2.64] & 1.20 [0.64] \\
  & SGLang      & 18371 [1.61] & 413.7 [3.35] & 1.10 [0.59] \\
\midrule
\multirow{4}{*}{\shortstack[l]{Qwen-3\\30B-A3B\\$\lambda \le 4$}}
  & \systemname & \textbf{1589.7 [1.14]} & \textbf{34.4 [0.97]} & \textbf{4.81 [0.99]} \\
  & TRT-LLM     & 23587 [4.90] & 604.9 [9.19] & 1.01 [0.28] \\
  & vLLM        & 17989 [2.02] & 276.8 [3.04] & 1.57 [0.54] \\
  & SGLang      & 23449 [1.98] & 478.1 [3.96] & 1.18 [0.45] \\
\end{tabular}%
}
\end{table}

\smartparagraph{\systemname's operating range survives interference intact.}
Across the full \systemname-defined operating range, \systemname remains effectively
unchanged under interference: TTFT inflation stays within
\numrange{0.92}{1.14}$\times$, TPOT inflation within
\numrange{0.97}{1.04}$\times$, and throughput at the saturation point within
\numrange{0.99}{1.02}$\times$ of isolation, showing that host CPU contention
does not perturb the steady-state decode path.

\begin{figure*}[t]
    \centering
    \raisebox{16mm}{\rotatebox{90}{\small Throughput (req/s)}}%
    \begin{subfigure}[b]{0.240\textwidth}
        \centering
        \includegraphics[width=\linewidth]{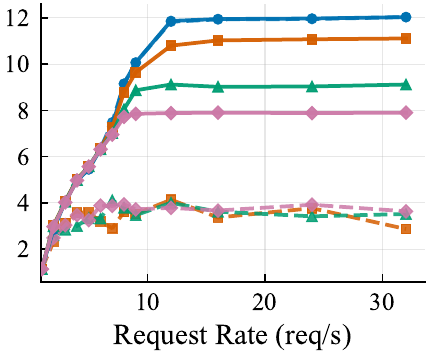}
        \caption{Llama-3 8B.}
    \end{subfigure}\hfill
    \begin{subfigure}[b]{0.240\textwidth}
        \centering
        \includegraphics[width=\linewidth]{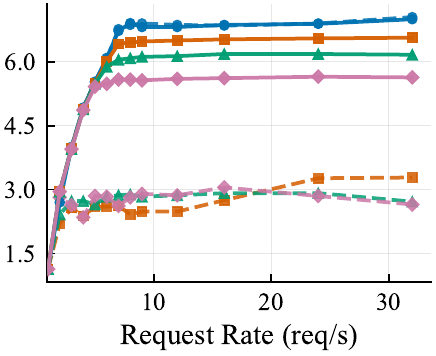}
        \caption{Phi-4 15B.}
    \end{subfigure}\hfill
    \begin{subfigure}[b]{0.240\textwidth}
        \centering
        \includegraphics[width=\linewidth]{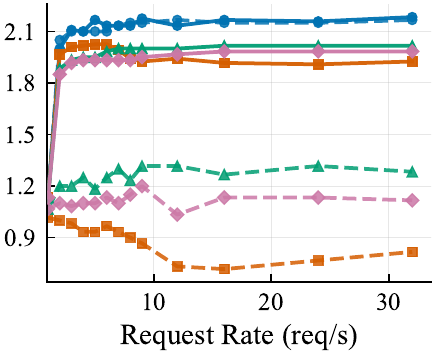}
        \caption{Qwen-3 32B.}
    \end{subfigure}\hfill
    \begin{subfigure}[b]{0.240\textwidth}
        \centering
        \includegraphics[width=\linewidth]{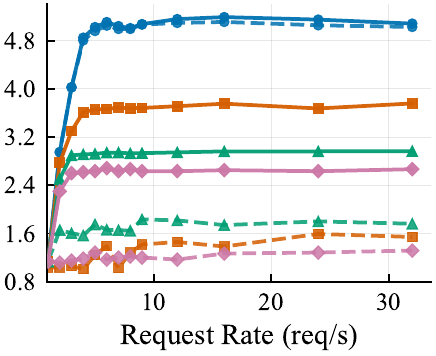}
        \caption{Qwen-3 30B-A3B.}
    \end{subfigure}
    \figlegend
    \caption{Throughput across four models. Solid lines show isolated execution; dashed lines show CPU interference. \systemname reaches the latest or tied-latest saturation point in isolation and preserves its plateau under interference, while baselines either saturate earlier or collapse to much lower plateaus.}
    \label{fig:throughput}
    \Description[Throughput under isolation and interference for all four models]{Throughput under isolation and interference for all four models.}
\end{figure*}

\smartparagraph{CPU-coupled baselines incur a large interference tax inside \systemname's range.}
Once we evaluate the same load ranges on CPU-coupled baselines, the contrast
is sharp.  On Llama-3 8B, baselines suffer \numrange{8.43}{18.84}$\times$
TTFT inflation, \numrange{5.77}{11.10}$\times$ TPOT inflation, and retain only
\numrange{0.38}{0.48}$\times$ of their isolated throughput at saturation.  On
Phi-4 15B, the corresponding inflation is \numrange{3.82}{10.66}$\times$ for
TTFT, \numrange{3.15}{6.17}$\times$ for TPOT, and
\numrange{0.41}{0.47}$\times$ for throughput retention.  Even on the GPU-bound
Qwen-3 32B, where isolated gaps are smallest, interference still inflates
TTFT by \numrange{1.54}{1.68}$\times$ and TPOT by
\numrange{2.64}{3.35}$\times$, while reducing throughput at saturation to
\numrange{0.51}{0.64}$\times$ of isolation.  The \acrshort{MoE} model shows the
clearest failure mode: within \systemname's operating range, TensorRT-LLM
incurs \num{4.90}$\times$ TTFT inflation and \num{9.19}$\times$ TPOT
inflation, while its throughput at saturation collapses from \qty{3.61}{req/s} to
\qty{1.01}{req/s}.  This behavior is visible directly in
Figs.~\ref{fig:p99-latency}(a--d),~\ref{fig:p99-latency}(e--h),
and~\ref{fig:throughput}(a--d), where the baseline dashed curves separate
sharply from their isolated counterparts.

\smartparagraph{After saturation, \systemname's plateau is preserved while baselines collapse.}
Fig.~\ref{fig:throughput}(a--d) shows that \systemname's saturation point is
unchanged under interference on all four models, and its plateau throughput is
preserved
(\qtyrange{99}{100}{\percent} retention).  In contrast, baseline plateau
retention falls to \qtyrange{32}{48}{\percent} on Llama-3 8B,
\qtyrange{42}{50}{\percent} on Phi-4 15B, \qtyrange{45}{64}{\percent} on
Qwen-3 32B, and \qtyrange{36}{59}{\percent} on Qwen-3 30B-A3B.  Under
interference, \systemname's plateau remains \numrange{3.18}{3.38}$\times$ higher
than the baselines on Llama-3 8B, \numrange{2.41}{2.50}$\times$ higher on
Phi-4 15B, \numrange{1.69}{2.42}$\times$ higher on Qwen-3 32B, and
\numrange{2.91}{4.08}$\times$ higher on Qwen-3 30B-A3B.

\subsection{Energy Efficiency (Q3)}
\label{sec:eval-energy}

We measure server-level wall power using a calibrated smart meter that records cumulative energy at \qty{1}{min} intervals from server's PSU feed. For \systemname, we additionally account for the BlueField-3 DPU's power consumption, sampled at \qty{0.5}{s} intervals via the DPU's on-board meter. We report energy per token (mJ/tok) as the product of average wall power and benchmark duration, divided by the number of tokens successfully processed.

\begin{figure}[t]
    \centering
    \includegraphics[width=\columnwidth]{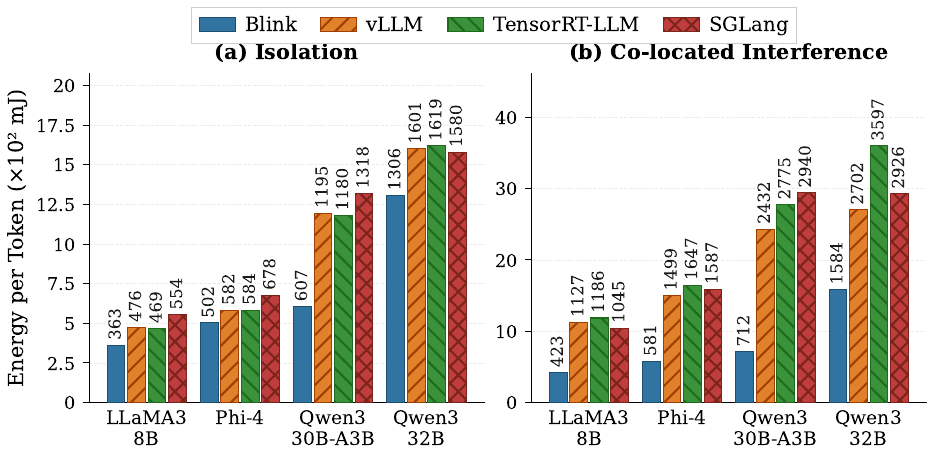}
    \caption{Energy per token under (a)~isolation and (b)~colocated interference.}
    \label{fig:energy}
    \Description[Energy per token for four models under isolation and interference]{Bar chart comparing energy per token (mJ/tok) for \systemname, vLLM, TensorRT-LLM, and SGLang across LLaMA 3 8B, Phi-4, Qwen3 30B-A3B, and Qwen3 32B under isolation and co-located interference.}
\end{figure}

\smartparagraph{Isolation.}
\systemname consumes \qtyrange{363}{1306}{mJ/tok} across four models (Fig.~\ref{fig:energy}(a)), \qtyrange{13.7}{48.6}{\percent} less than the most efficient baseline per model.
The gap ranges from Phi-4 (\num{502} vs.\ \qty{582}{mJ/tok} for vLLM, where model size limits relative scheduling overhead) to Qwen-3 30B-A3B (\num{607} vs.\ \qty{1180}{mJ/tok} for TensorRT-LLM, where \acrshort{MoE} expert routing amplifies CPU scheduling overhead).
For Qwen-3 32B, the largest dense model, \systemname achieves \qty{1306}{mJ/tok} versus \qty{1580}{mJ/tok} for SGLang---a \qty{17.3}{\percent} reduction.

\smartparagraph{Interference.}
Under colocated interference (Fig.~\ref{fig:energy}(b)), \systemname\linebreak[4] achieves \qtyrange{423}{1584}{mJ/tok} while CPU-mediated baselines rise to \qtyrange{1045}{3597}{mJ/tok}, a \qtyrange{41.4}{70.7}{\percent} reduction relative to the best baseline per model.
All four systems draw comparable wall power (\qtyrange{1.1}{1.4}{kW}), so energy per token tracks inversely with throughput.
When CPU contention collapses baseline throughput (\S\ref{sec:eval-interference}) at constant power, their energy per token inflates by \qtyrange{69}{182}{\percent}.
\systemname's overhead is at most \qty{21}{\percent}: removing the host CPU from the critical path makes energy efficiency structurally independent of host contention.

\section{Discussion}
\label{sec:discussion}
A natural concern is whether removing the host CPU precludes standard serving optimizations; we argue it does not, as \systemname exposes clear extension points for each major technique.

\smartparagraph{Tensor parallelism and pipeline parallelism.}
While \systemname's current prototype targets a single GPU, the same control-plane structure extends to multi-GPU deployments. Tensor-parallel sharding and pipeline parallelism can be realized by instantiating a persistent scheduler on each GPU (with its own local shared-memory region) and inserting GPU-native communication primitives (\eg NCCL collectives or GPU-initiated RDMA via IBGDA~\cite{deepep}) between graph executions; device-side synchronization enforces the required ordering. For \acrshort{MoE} models, GPU-initiated all-to-all dispatch~\cite{deepep} is particularly attractive, as it avoids NCCL's CPU proxy overhead. Because these mechanisms execute on the GPUs, they preserve \systemname's CPU-free critical path.
A full implementation and evaluation of these multi-GPU extensions is left to future work.

\smartparagraph{Serving optimizations.}
Several widely studied optimizations map naturally onto \systemname's GPU-resident scheduler. \emph{Chunked prefill}~\cite{sarathi_serve}---the scheduler already tracks per-request progress and KV-cache status, enabling incremental prefill without changing the data plane.
\emph{Prefix caching}~\cite{sglang}---the paged KV cache provides reusable blocks; a GPU-resident trie or hash table can map token prefixes to KV-block ranges inside the scheduler.
\emph{Speculative decoding}~\cite{spec-decoding}---the scheduler runs draft generation followed by target verification and publishes only accepted tokens to the ring buffer.
\emph{Disaggregated prefill/decode}~\cite{distserve}---separate scheduler instances for prefill and decode can transfer KV state over NVLink without involving the CPU.
\emph{KV cache offloading}---when HBM is exhausted, inactive KV blocks can be offloaded to a peer GPU via NVLink-based device-initiated transfers~\cite{aqua} or CUDA managed memory.

\smartparagraph{CPU/DRAM offloading.} When models exceed GPU memory capacity, which is increasingly common with \gls{MoE} architectures whose total parameter counts far exceed their active parameters, systems resort to CPU/DRAM offloading for expert weights~\cite{pregated_moe, huang2025moe_offload}. In this regime the host CPU is not only on the scheduling path but also on the data-movement path: it must orchestrate dynamic expert routing and migrate weights between host memory and GPU memory on every token. This makes the interference problem described in \S\ref{sec:characterization} strictly worse, as expert migration latency compounds with orchestration jitter and memory bus contention directly stalls expert fetches. \systemname currently targets the common case where the served model fits in GPU memory. Extending the architecture to offloading scenarios is a natural next step; the DPU's RDMA engine could manage expert migration directly, keeping weight transfers off the host memory bus while preserving the CPU-free serving path.

\smartparagraph{Broader ecosystem trends.}
\systemname's ARM-based DPU frontend is well-aligned with the growing role of ARM in AI datacenter infrastructure, exemplified by the Arm AGI CPU~\cite{arm_agi_cpu} co-developed with Meta. As ARM-based DPUs gain higher core counts and tighter RDMA integration, \systemname's frontend stands to benefit directly.

\section{Related Work}
\label{sec:related-work}
\smartparagraph{SmartNIC/DPU offload for accelerator services.}
The most closely related systems offload portions of ML or storage datapaths to SmartNICs, but all target either one-shot DNN inference or non-inference workloads.
Lynx~\cite{lynx} proposes an accelerator-centric architecture where the SmartNIC replaces the host CPU as the orchestrator for GPU network services: it receives requests, dispatches them to GPUs via RDMA, and returns responses without host involvement. It operates at request-level granularity on one-shot workloads (face verification, LeNet) with no support for autoregressive and stateful LLMs.
SplitRPC~\cite{10.1145/3589974} steers inference payloads directly from the NIC to GPU memory, reducing host-side RPC data-movement overheads, yet retains the host CPU for  orchestration, limiting it to stateless, one-shot DNN inference.
Conspirator~\cite{298581} moves an ML control plane onto SmartNIC ARM cores for job-level scheduling across GPU instances in distributed training, but the host CPU remains on the critical path for kernel invocation and result retrieval, and scheduling operates at job granularity rather than per-token.
OS2G~\cite{os2g} offloads an object-storage client onto a DPU, introducing GPUDirect DPU for direct DPU-to-GPU transfers and fully bypassing the host for deep-learning training I/O, validating the same DPU-to-GPU transfer primitive \systemname uses, but targeting storage rather than inference. DeepEP~\cite{deepep} uses GPU-initiated RDMA (IBGDA) to bypass NCCL's CPU proxy for \acrshort{MoE} expert-parallel dispatch, but targets a single collective primitive within host-driven stacks rather than end-to-end serving.

\smartparagraph{LLM serving and GPU execution}
A large body of work optimizes the host-driven serving loop: continuous batching~\cite{orca}, chunked prefills~\cite{sarathi_serve}, prefill/decode disaggregation~\cite{distserve}, PagedAttention~\cite{vllm}, and further scheduling and memory optimizations~\cite{fastserve,vattention,flexgen,deepspeed_inference}. On the GPU side, CUDA Graphs~\cite{cuda_graphs}, device-side graph launch~\cite{cuda_device_graph_launch}, memory-efficient attention~\cite{flashattention,flashattention2}, and production runtimes such as TensorRT-LLM~\cite{tensorrt_llm} accelerate individual operations. All retain the host CPU as the locus of scheduling and I/O; \systemname instead runs these batching and memory policies in its GPU-resident scheduler, keeping the host off the steady-state critical path.


\vspace{.05in}
\noindent None of these systems addresses the challenges unique to autoregressive LLM inference: token-level continuous batching, stateful KV-cache management across long-lived decode sessions, and per-token streaming. \systemname is the first to restructure the inference serving stack end to end, to eliminate the host CPU from the entire autoregressive inference lifecycle.

\section{Conclusion}
\label{sec:conclusio}

This paper introduced \systemname, a CPU-free LLM serving architecture that removes the host CPU from the steady-state inference path by co-designing SmartNIC-resident ingress, zero-copy RDMA into GPU memory, and a GPU-resident persistent scheduler for continuous batching, KV-cache management, and token generation. 
Across four models spanning dense and \acrshort{MoE} architectures (Llama-3 8B, Phi-4 15B, Qwen-3 32B, and Qwen-3 30B-A3B), \systemname improves pre-saturation P99 TTFT by up to \num{8.47}$\times$ and P99 TPOT by up to \num{3.40}$\times$, decode throughput by up to \num{2.1}$\times$, and energy per token by up to \qty{48.6}{\percent} over TensorRT-LLM, vLLM, and SGLang, while maintaining performance under multi-tenant CPU contention where baselines degrade by one to two orders of magnitude.

More broadly, \systemname shows that persistent accelerator-resident control combined with SmartNIC-resident I/O can replace host-mediated orchestration for latency-sensitive workloads, and that decoupling inference from shared host resources will be key to performance isolation and  server consolidation as datacenters scale.


\bibliographystyle{ACM-Reference-Format}
\bibliography{references}

\appendix
\counterwithin{figure}{section}
\counterwithin{table}{section}

\clearpage
\noindent{\Large\textbf{Appendix}}
\vspace{1em}

\noindent
This appendix provides supplementary evaluation results for \systemname.
The main paper presents P99 TTFT, P99 TPOT, and goodput under both
isolated and colocated conditions; here we extend the analysis to
additional percentiles (P99.9, P95, P50, Mean), inter-token latency
(ITL), and token-level throughput (prefill and decode tokens/s).
All figures show both isolated and CPU-interference conditions across
all four models: Llama-3 8B, Phi-4 15B, Qwen-3 32B, and Qwen-3 30B-A3B.
We also include summary tables with geometric-mean latency and
throughput comparisons over \systemname's operating range.

\section{Experimental Configuration}
\label{sec:app-setup}

We briefly recap the experimental setup; full details are in the main
paper (\S\ref{sec:evaluation}).  All experiments run on a single NVIDIA H100
(96\,GB HBM3) server with two Intel Xeon Gold 6336Y CPUs (96 cores),
256\,GB DDR5, and a ConnectX-6 200\,Gbps NIC.  \systemname's frontend runs
on a BlueField-3 DPU connected via RDMA.

We compare four systems: \systemname, TensorRT-LLM v1.1.0 (using the
same TensorRT v10.14 engines as \systemname), vLLM v0.13.0, and SGLang
v0.5.8.  The workload uses the ShareGPT v3 dataset with 13 offered-load
levels from 1 to 32\,req/s.  The interference workload colocates
\textit{pbzip2} (45 threads) and a \textit{Ninja} LLVM build (45 jobs)
on the remaining 90 host cores.
Both interference workloads are launched once before the sweep begins and
run continuously throughout all 13 load levels.  Because these workloads
traverse distinct execution phases over their lifetime (Ninja cycles
through preprocessing, compilation, assembly, and linking; pbzip2
alternates between I/O-intensive reads and compute-intensive compression
blocks), the resource interference they impose varies over time.  Consequently, each rate point in the sweep encounters a
different interference profile, which can produce non-monotonic
fluctuations in baseline interference curves across successive offered
loads.

The four models span a range of sizes and architectures:
Llama-3 8B (dense, 8B parameters),
Phi-4 15B (dense, 14.7B),
Qwen-3 32B (dense, 32B), and
Qwen-3 30B-A3B (MoE, 30B total / 3B active).
In all figures, solid lines denote isolated execution and dashed lines
denote execution under CPU interference.

\section{Pre-Saturation Latency Summary}
\label{sec:app-summary}

Table~\ref{tab:latency-summary} provides a comprehensive latency
comparison across all percentiles.  For each model, we report the
geometric mean over \systemname's operating range
($\lambda \le 12$ for Llama-3 8B, $\lambda \le 7$ for Phi-4 15B,
$\lambda \le 2$ for Qwen-3 32B, $\lambda \le 4$ for Qwen-3 30B-A3B)
under isolated execution.  The geometric mean is used (consistent with
the main paper) because it is less sensitive to a single high-load
outlier and is appropriate for ratio-scale latency data.

\smartparagraph{P50 TTFT.}
\systemname achieves the lowest median TTFT on all models.  On Llama-3 8B,
baselines incur 1.73$\times$ (TensorRT-LLM, \qty{72.3}{ms}) to
5.75$\times$ (SGLang, \qty{240.3}{ms}) higher median TTFT than
\systemname (\qty{41.8}{ms}).  The Qwen-3 30B-A3B result is notable:
TensorRT-LLM's \qty{1132}{ms} median TTFT is 5.5$\times$ higher than
\systemname's \qty{207}{ms}, reflecting the high cost of CPU-mediated
expert routing at even the median.

\smartparagraph{Mean latency.}
Mean TTFT and TPOT track between P50 and P95, as expected for
heavy-tailed queuing distributions.  On Phi-4 15B, \systemname achieves
a mean TTFT of \qty{258.8}{ms} compared to \qty{355.7}{ms} for
TensorRT-LLM (1.37$\times$) and \qty{846.9}{ms} for SGLang
(3.27$\times$).  Mean TPOT differences are smaller but consistent:
\systemname's per-token decode overhead is the lowest or near-lowest on
every model.

\smartparagraph{Qwen-3 32B: the GPU-bound regime.}
On the largest dense model, differences between systems are smallest
across all percentiles.  This is consistent with the main paper's finding
that once the workload becomes predominantly GPU-bound, there is less
host-side latency to remove.  Nevertheless, \systemname maintains parity
or a slight edge, and the advantage re-emerges at P99.9
(\S\ref{sec:app-p999} below).

\begin{table}[t]
\centering
\caption{Geometric-mean latency (ms) over \systemname's operating range
under isolated execution.  Best values per model are boldfaced.
Operating ranges: $\lambda \le 12$ (Llama-3 8B), $\le 7$ (Phi-4 15B),
$\le 2$ (Qwen-3 32B), $\le 4$ (Qwen-3 30B-A3B).}
\label{tab:latency-summary}
\resizebox{\linewidth}{!}{%
\begin{tabular}{p{1.8cm}lrrrr}
\textbf{Model} & \textbf{System}
  & \shortstack[c]{\textbf{P50}\\\textbf{TTFT}}
  & \shortstack[c]{\textbf{Mean}\\\textbf{TTFT}}
  & \shortstack[c]{\textbf{P50}\\\textbf{TPOT}}
  & \shortstack[c]{\textbf{Mean}\\\textbf{TPOT}} \\
\midrule
\multirow{4}{*}{Llama-3 8B}
  & \systemname & \textbf{41.8}  & \textbf{116.9} & \textbf{7.5}  & \textbf{8.2}  \\
  & TRT-LLM     & 72.3  & 170.1 & 8.6  & 9.6  \\
  & vLLM        & 111.1 & 276.1 & 11.0 & 12.4 \\
  & SGLang      & 240.3 & 507.1 & 12.9 & 14.5 \\
\midrule
\multirow{4}{*}{Phi-4 15B}
  & \systemname & \textbf{105.8} & \textbf{258.8} & \textbf{13.4} & \textbf{14.1} \\
  & TRT-LLM     & 153.1 & 355.7 & 15.2 & 16.3 \\
  & vLLM        & 229.5 & 455.8 & 16.8 & 18.1 \\
  & SGLang      & 540.9 & 846.9 & 20.6 & 22.4 \\
\midrule
\multirow{4}{*}{Qwen-3 32B}
  & \systemname & 786.2 & 2501  & \textbf{29.7} & \textbf{35.9} \\
  & TRT-LLM     & \textbf{531.7} & \textbf{2344} & 31.0 & 37.2 \\
  & vLLM        & 756.0 & 3009  & 32.7 & 40.8 \\
  & SGLang      & 788.5 & 2896  & 34.2 & 40.9 \\
\midrule
\multirow{4}{*}{\shortstack[l]{Qwen-3\\30B-A3B}}
  & \systemname & \textbf{207.1} & \textbf{426.1} & \textbf{11.9} & \textbf{13.8} \\
  & TRT-LLM     & 1132  & 1405  & 21.3 & 24.3 \\
  & vLLM        & 1507  & 2620  & 25.6 & 30.9 \\
  & SGLang      & 1778  & 3441  & 28.4 & 35.2 \\
\end{tabular}%
}
\end{table}

Table~\ref{tab:throughput-summary} shows token-level throughput at each
model's saturation point.  Decode throughput is the most
scheduling-sensitive metric: each autoregressive step requires a
scheduler iteration, so CPU-mediated overhead compounds linearly with
output length.  \systemname achieves \qty{3880}{tok/s} decode throughput
on Llama-3 8B at saturation (10\% above TensorRT-LLM) and
\qty{1437}{tok/s} on Qwen-3 30B-A3B (36\% above TensorRT-LLM).

\begin{table}[t]
\centering
\caption{Token-level throughput (tok/s) at \systemname's saturation
point under isolated execution.  Best values per model are boldfaced.}
\label{tab:throughput-summary}
\resizebox{0.9\linewidth}{!}{%
\begin{tabular}{p{1.8cm}lrr}
\textbf{Model} & \textbf{System}
  & \shortstack[c]{\textbf{Decode}\\\textbf{tok/s}}
  & \shortstack[c]{\textbf{Prefill}\\\textbf{tok/s}} \\
\midrule
\multirow{4}{*}{Llama-3 8B}
  & \systemname & \textbf{3880} & \textbf{595} \\
  & TRT-LLM     & 3535 & 582 \\
  & vLLM        & 2930 & 564 \\
  & SGLang      & 2638 & 553 \\
\midrule
\multirow{4}{*}{Phi-4 15B}
  & \systemname & \textbf{2177} & \textbf{465} \\
  & TRT-LLM     & 2044 & 459 \\
  & vLLM        & 1906 & 451 \\
  & SGLang      & 1690 & 444 \\
\midrule
\multirow{4}{*}{Qwen-3 32B}
  & \systemname & \textbf{537} & \textbf{236} \\
  & TRT-LLM     & 520 & 235 \\
  & vLLM        & 487 & 229 \\
  & SGLang      & 482 & 229 \\
\midrule
\multirow{4}{*}{\shortstack[l]{Qwen-3\\30B-A3B}}
  & \systemname & \textbf{1437} & \textbf{374} \\
  & TRT-LLM     & 1053 & 318 \\
  & vLLM        & 841  & 262 \\
  & SGLang      & 730  & 257 \\
\end{tabular}%
}
\end{table}

\section{Serviceable Load Summary}
\label{sec:app-serviceable}

\begin{figure}[!t]
    \centering
    \captionsetup[subfigure]{}
    \begin{subfigure}[t]{0.8\columnwidth}
        \centering
        \includegraphics[width=\columnwidth]{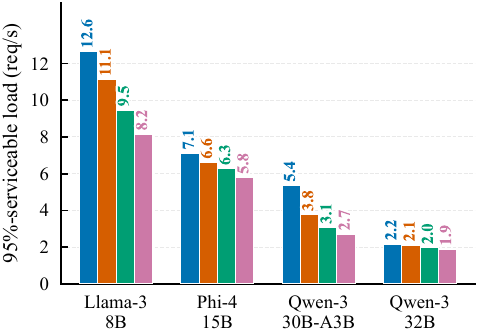}
        \caption{Isolated.}
        \label{fig:serviceable-load-isolated}
    \end{subfigure}

    \begin{subfigure}[t]{0.8\columnwidth}
        \centering
        \includegraphics[width=\columnwidth]{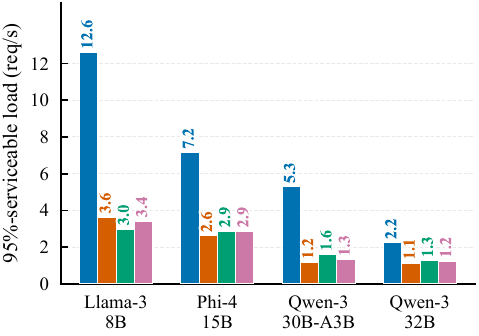}
        \caption{Interference.}
        \label{fig:serviceable-load-interference}
    \end{subfigure}

    \barlegend

    \caption{Maximum serviceable load (95\% throughput retention threshold) per system and model. Higher is better. \systemname achieves the highest serviceable load on every model; on Qwen-3 30B-A3B the gap is largest under isolation, and under interference \systemname retains its full isolated capacity while baselines lose 60--65\% on Llama-3 8B.}
    \label{fig:serviceable-load}
    \Description[Serviceable load bar charts under isolated and interference]{Two bar charts showing maximum serviceable load for all systems across four models: (a) isolated and (b) interference conditions.}
\end{figure}

Fig.~\ref{fig:serviceable-load} summarizes the maximum serviceable
load for each system-model pair under isolated
(\subref{fig:serviceable-load-isolated}) and interference
(\subref{fig:serviceable-load-interference}) conditions.  The serviceable load is defined as
the highest offered rate at which the system retains at least 95\% of
the ideal throughput (\ie goodput $\geq$ 0.95 $\times$ offered rate).
This metric captures the practical deployment capacity: the maximum
request rate at which a system can operate without significant queuing or
request dropping.

\systemname achieves the highest serviceable load on every model under
both isolated and interference conditions.  On the MoE model
(Qwen-3 30B-A3B), the gap is particularly large: \systemname sustains
nearly twice the serviceable load of the nearest baseline under
isolation, and the advantage widens further under interference where
CPU-mediated systems lose scheduling capacity to the competing workloads.
On Llama-3 8B, the serviceable load under interference drops by
60--65\% for baselines while \systemname retains its full isolated
capacity.

\section{Extended Latency Metrics}
\label{sec:app-latency}

The main paper reports P99 tail latency for TTFT and TPOT (Fig.~6).
Here we present four additional percentile breakdowns: P99.9, P95,
P50 (median), and Mean.  Each figure covers all four models under both
isolated and interference conditions.  P99.9 and P95 figures show TTFT
and TPOT; P50 and Mean figures also include ITL.

These percentiles serve complementary purposes.
P99.9 exposes outlier behavior at the deepest practical tail; P95
captures the experience of the vast majority of requests without the
noise sensitivity of P99.9; P50 (median) reflects the typical-case
latency; and the mean provides an aggregate view useful for capacity
planning.  ITL (inter-token latency) is relevant for streaming
applications where users perceive token-by-token delivery smoothness.

\subsection{P99.9 Tail Latency}
\label{sec:app-p999}

\begin{figure*}[!t]
    \centering
    \raisebox{16mm}{\rotatebox{90}{\small P99.9 TTFT (ms)}}%
    \begin{subfigure}[b]{0.240\textwidth}
        \centering
        \includegraphics[width=\linewidth]{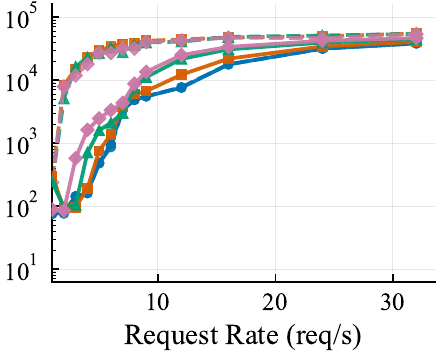}
        \caption{Llama-3 8B}
    \end{subfigure}\hfill
    \begin{subfigure}[b]{0.240\textwidth}
        \centering
        \includegraphics[width=\linewidth]{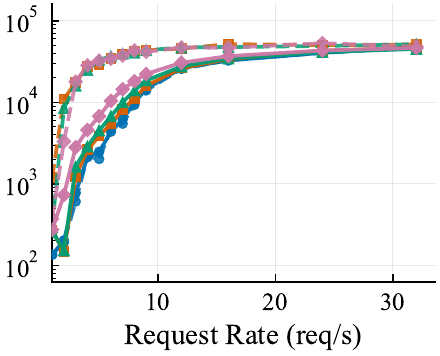}
        \caption{Phi-4 15B}
    \end{subfigure}\hfill
    \begin{subfigure}[b]{0.240\textwidth}
        \centering
        \includegraphics[width=\linewidth]{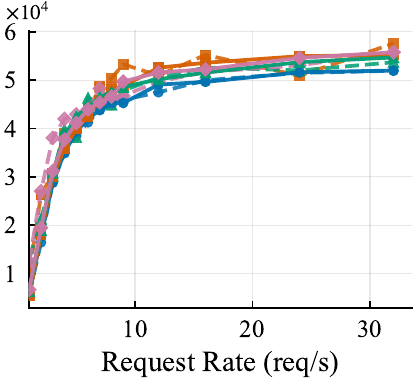}
        \caption{Qwen-3 32B}
    \end{subfigure}\hfill
    \begin{subfigure}[b]{0.240\textwidth}
        \centering
        \includegraphics[width=\linewidth]{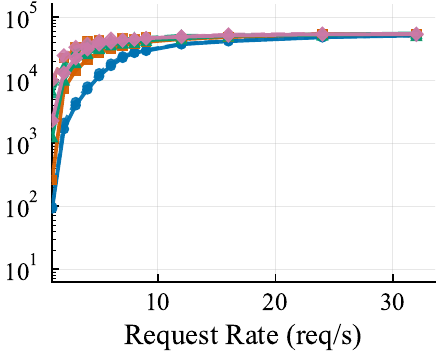}
        \caption{Qwen-3 30B-A3B}
    \end{subfigure}\\[2pt]
    \raisebox{16mm}{\rotatebox{90}{\small P99.9 TPOT (ms)}}%
    \begin{subfigure}[b]{0.240\textwidth}
        \centering
        \includegraphics[width=\linewidth]{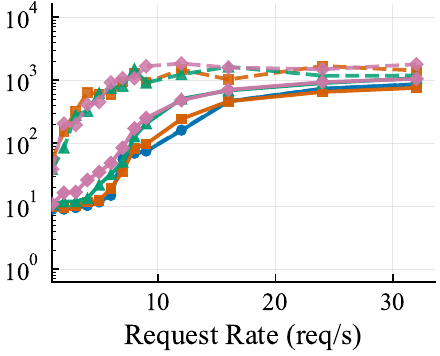}
        \caption{Llama-3 8B}
    \end{subfigure}\hfill
    \begin{subfigure}[b]{0.240\textwidth}
        \centering
        \includegraphics[width=\linewidth]{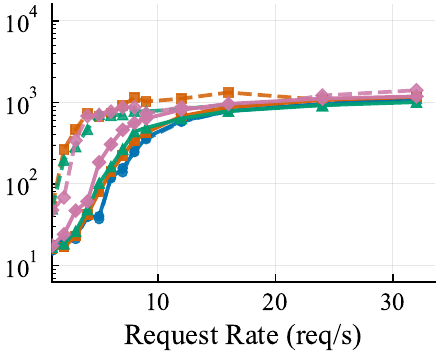}
        \caption{Phi-4 15B}
    \end{subfigure}\hfill
    \begin{subfigure}[b]{0.240\textwidth}
        \centering
        \includegraphics[width=\linewidth]{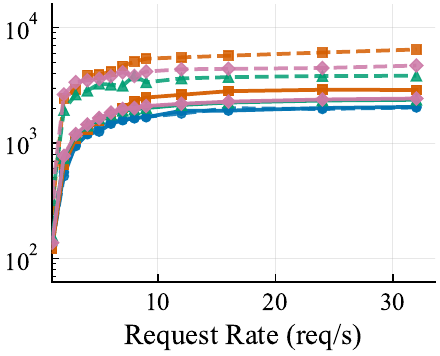}
        \caption{Qwen-3 32B}
    \end{subfigure}\hfill
    \begin{subfigure}[b]{0.240\textwidth}
        \centering
        \includegraphics[width=\linewidth]{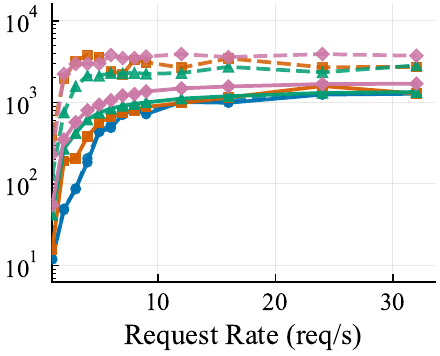}
        \caption{Qwen-3 30B-A3B}
    \end{subfigure}
    \figlegend
    \caption{P99.9 tail latency across four models.  Top row~(a--d): TTFT; bottom row~(e--h): TPOT.  Solid lines show isolated execution; dashed lines show CPU interference.  \systemname sustains lower P99.9 latency across the full load sweep on all four models, with the largest margin on the MoE model (Qwen-3 30B-A3B).  Baseline dashed curves separate sharply from their isolated counterparts, while \systemname's remain nearly overlapping.}
    \label{fig:p999-all}
    \Description[P99.9 latency for all models]{P99.9 TTFT and TPOT across four models under isolated and interference conditions.}
\end{figure*}

Fig.~\ref{fig:p999-all} extends the P99.9 analysis from the main
paper to all four models.

\smartparagraph{Isolation.}
At this deep tail, \systemname consistently maintains lower latency
across the full load range.  The separation is most pronounced on
Qwen-3 30B-A3B, where MoE expert-routing overhead in CPU-mediated
systems amplifies tail events.  On Llama-3 8B, baselines incur
1.3--2.6$\times$ higher P99.9 TTFT than \systemname within the operating
range, consistent with the P99 ratios reported in the main paper.

\smartparagraph{Interference.}
Under CPU contention, baseline P99.9 latencies diverge sharply from
their isolated counterparts (dashed vs.\ solid curves), while
\systemname's curves remain nearly overlapping.  This confirms that the
interference immunity documented at P99 in the main paper extends to the
deepest practical tail.  The effect is especially visible in the TPOT row
(e--h), where per-iteration scheduling delays compound across hundreds
of decode steps.

\subsection{P95 Tail Latency}
\label{sec:app-p95}

\begin{figure*}[!t]
    \centering
    \raisebox{16mm}{\rotatebox{90}{\small P95 TTFT (ms)}}%
    \begin{subfigure}[b]{0.240\textwidth}
        \centering
        \includegraphics[width=\linewidth]{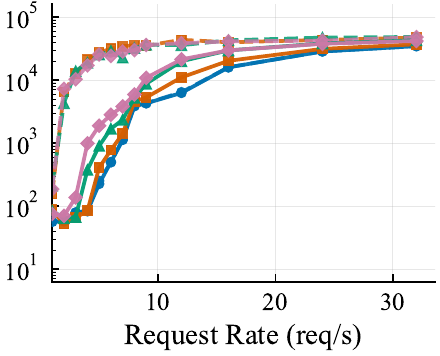}
        \caption{Llama-3 8B}
    \end{subfigure}\hfill
    \begin{subfigure}[b]{0.240\textwidth}
        \centering
        \includegraphics[width=\linewidth]{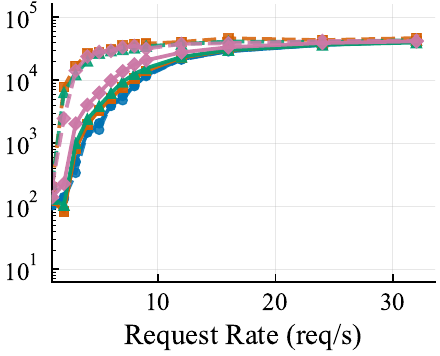}
        \caption{Phi-4 15B}
    \end{subfigure}\hfill
    \begin{subfigure}[b]{0.240\textwidth}
        \centering
        \includegraphics[width=\linewidth]{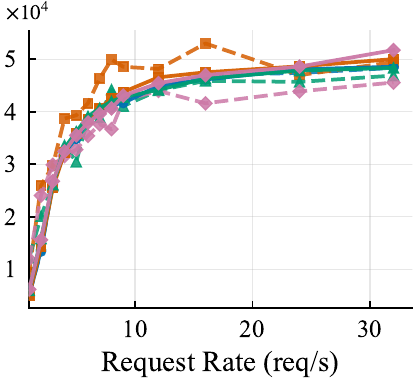}
        \caption{Qwen-3 32B}
    \end{subfigure}\hfill
    \begin{subfigure}[b]{0.240\textwidth}
        \centering
        \includegraphics[width=\linewidth]{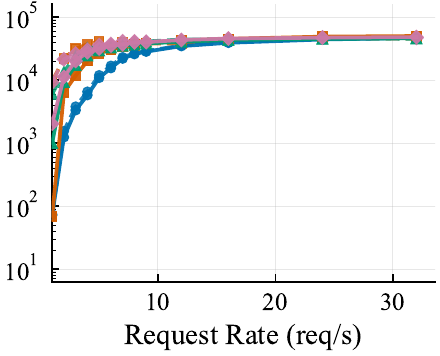}
        \caption{Qwen-3 30B-A3B}
    \end{subfigure}\\[2pt]
    \raisebox{16mm}{\rotatebox{90}{\small P95 TPOT (ms)}}%
    \begin{subfigure}[b]{0.240\textwidth}
        \centering
        \includegraphics[width=\linewidth]{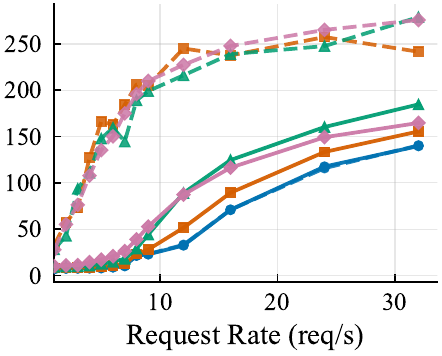}
        \caption{Llama-3 8B}
    \end{subfigure}\hfill
    \begin{subfigure}[b]{0.240\textwidth}
        \centering
        \includegraphics[width=\linewidth]{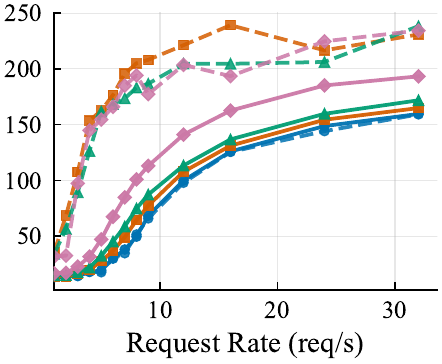}
        \caption{Phi-4 15B}
    \end{subfigure}\hfill
    \begin{subfigure}[b]{0.240\textwidth}
        \centering
        \includegraphics[width=\linewidth]{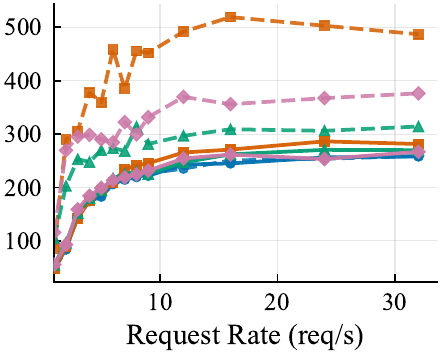}
        \caption{Qwen-3 32B}
    \end{subfigure}\hfill
    \begin{subfigure}[b]{0.240\textwidth}
        \centering
        \includegraphics[width=\linewidth]{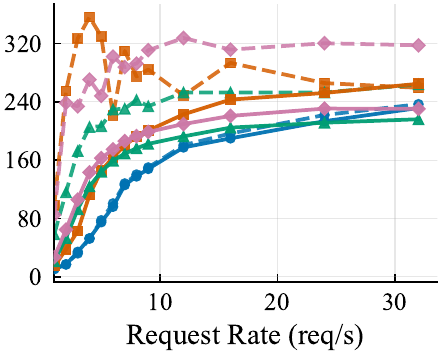}
        \caption{Qwen-3 30B-A3B}
    \end{subfigure}
    \figlegend
    \caption{P95 latency across four models.  Top row~(a--d): TTFT; bottom row~(e--h): TPOT.  Solid lines show isolated execution; dashed lines show CPU interference.  The relative system ordering is consistent with the P99 results in the main paper, confirming that \systemname's advantage is structural rather than an artifact of rare outliers.}
    \label{fig:p95-all}
    \Description[P95 latency for all models]{P95 TTFT and TPOT across four models under isolated and interference conditions.}
\end{figure*}

Fig.~\ref{fig:p95-all} reports P95 latency.  At this percentile, the
relative ordering among systems is consistent with P99 (main paper), but
absolute values are lower and the curves are smoother due to reduced
sensitivity to individual outlier requests.

\smartparagraph{Isolation.}
\systemname retains its latency advantage across all models.  The P95
TTFT gaps are proportionally similar to P99: on Llama-3 8B, baselines
incur 1.3--2.4$\times$ higher P95 TTFT than \systemname within the
operating range.  This confirms that the benefit is structural (reduced
scheduling overhead) rather than an artifact of rare tail events.

\smartparagraph{Interference.}
The separation between isolated and interference curves is visible at P95
for all CPU-mediated baselines, while \systemname's P95 curves remain
stable.  Under interference, baseline P95 TPOT rises steeply past
saturation, directly impacting per-token decode latency for interactive
applications.

\subsection{P50 (Median) Latency}
\label{sec:app-p50}

\begin{figure*}[!t]
    \centering
    \raisebox{16mm}{\rotatebox{90}{\small P50 TTFT (ms)}}%
    \begin{subfigure}[b]{0.240\textwidth}
        \centering
        \includegraphics[width=\linewidth]{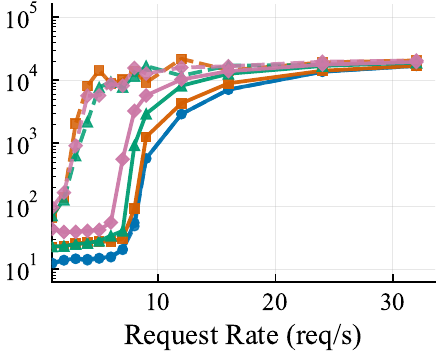}
        \caption{Llama-3 8B}
    \end{subfigure}\hfill
    \begin{subfigure}[b]{0.240\textwidth}
        \centering
        \includegraphics[width=\linewidth]{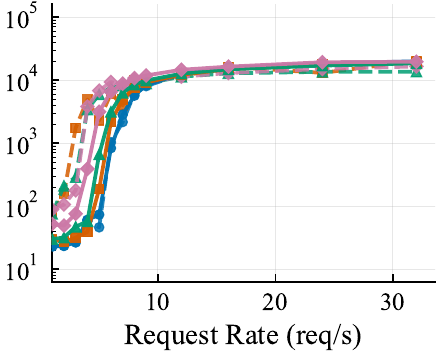}
        \caption{Phi-4 15B}
    \end{subfigure}\hfill
    \begin{subfigure}[b]{0.240\textwidth}
        \centering
        \includegraphics[width=\linewidth]{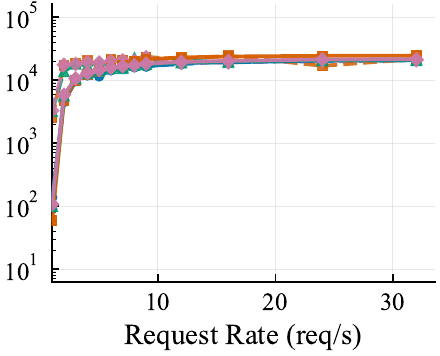}
        \caption{Qwen-3 32B}
    \end{subfigure}\hfill
    \begin{subfigure}[b]{0.240\textwidth}
        \centering
        \includegraphics[width=\linewidth]{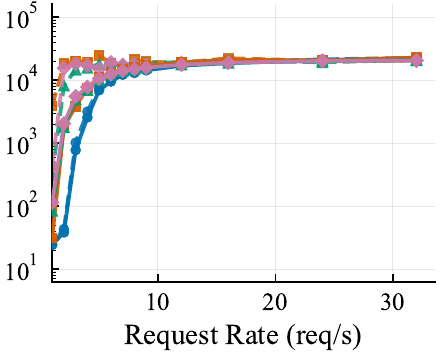}
        \caption{Qwen-3 30B-A3B}
    \end{subfigure}\\[2pt]
    \raisebox{16mm}{\rotatebox{90}{\small P50 TPOT (ms)}}%
    \begin{subfigure}[b]{0.240\textwidth}
        \centering
        \includegraphics[width=\linewidth]{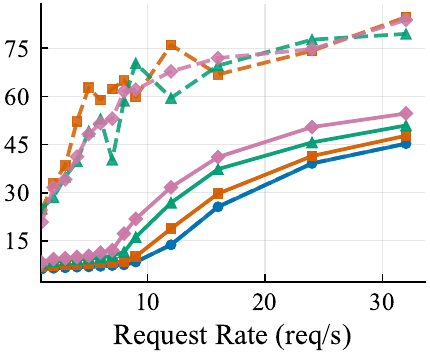}
        \caption{Llama-3 8B}
    \end{subfigure}\hfill
    \begin{subfigure}[b]{0.240\textwidth}
        \centering
        \includegraphics[width=\linewidth]{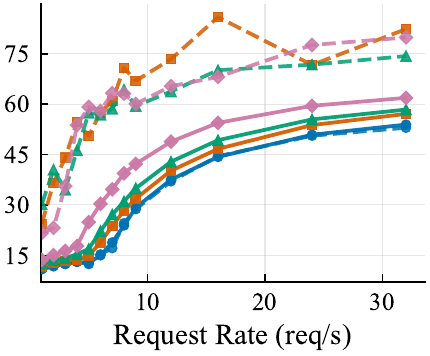}
        \caption{Phi-4 15B}
    \end{subfigure}\hfill
    \begin{subfigure}[b]{0.240\textwidth}
        \centering
        \includegraphics[width=\linewidth]{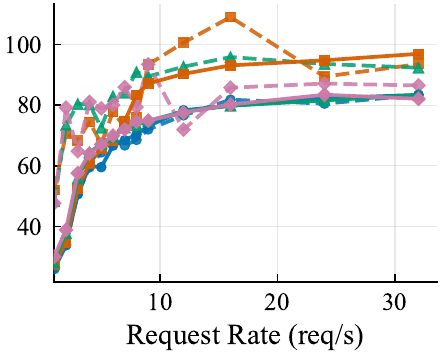}
        \caption{Qwen-3 32B}
    \end{subfigure}\hfill
    \begin{subfigure}[b]{0.240\textwidth}
        \centering
        \includegraphics[width=\linewidth]{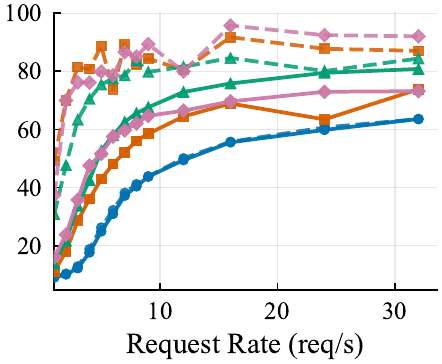}
        \caption{Qwen-3 30B-A3B}
    \end{subfigure}\\[2pt]
    \raisebox{16mm}{\rotatebox{90}{\small P50 ITL (ms)}}%
    \begin{subfigure}[b]{0.240\textwidth}
        \centering
        \includegraphics[width=\linewidth]{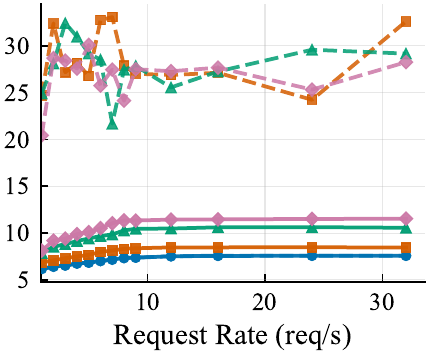}
        \caption{Llama-3 8B}
    \end{subfigure}\hfill
    \begin{subfigure}[b]{0.240\textwidth}
        \centering
        \includegraphics[width=\linewidth]{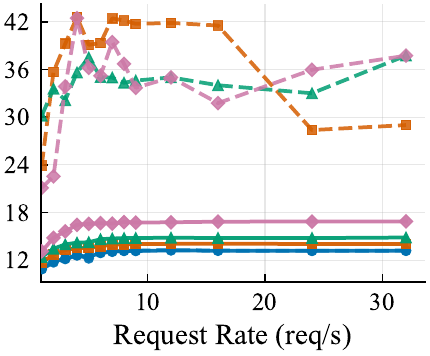}
        \caption{Phi-4 15B}
    \end{subfigure}\hfill
    \begin{subfigure}[b]{0.240\textwidth}
        \centering
        \includegraphics[width=\linewidth]{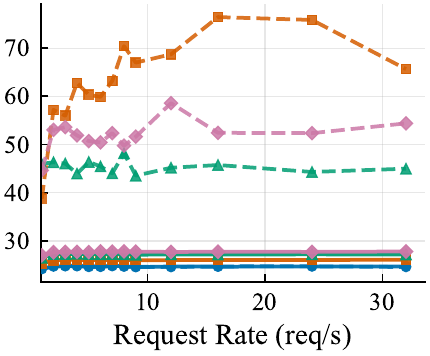}
        \caption{Qwen-3 32B}
    \end{subfigure}\hfill
    \begin{subfigure}[b]{0.240\textwidth}
        \centering
        \includegraphics[width=\linewidth]{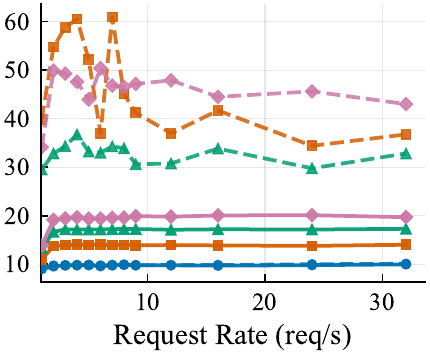}
        \caption{Qwen-3 30B-A3B}
    \end{subfigure}
    \figlegend
    \caption{Median (P50) latency across four models.  Top row~(a--d): TTFT; middle row~(e--h): TPOT; bottom row~(i--l): ITL.  Solid lines show isolated execution; dashed lines show CPU interference.  \systemname achieves the lowest or near-lowest median across all metrics, with differences compressing on the GPU-bound Qwen-3 32B.  Under interference, baseline medians inflate substantially while \systemname's remain stable.}
    \label{fig:p50-all}
    \Description[P50 latency for all models]{P50 TTFT, TPOT, and ITL across four models under isolated and interference conditions.}
\end{figure*}

Fig.~\ref{fig:p50-all} shows median latency.  Median values are
important for capacity planning because they reflect the experience of a
typical request rather than worst-case outliers.

\smartparagraph{Isolation.}
\systemname achieves the lowest or near-lowest median across all metrics
and models.  On Qwen-3 32B, where the workload is heavily GPU-bound,
median differences between systems are small
(Table~\ref{tab:latency-summary}), consistent with the P99 observations
in the main paper.  On Qwen-3 30B-A3B the picture is different:
TensorRT-LLM's median TTFT (\qty{1132}{ms}) is 5.5$\times$ higher than
\systemname's (\qty{207}{ms}), confirming that expert-routing
overhead is not confined to the tail.

\smartparagraph{Interference.}
At the median, CPU interference shifts baseline curves upward
significantly.  Baselines that were within 2$\times$ of \systemname in
isolation can exceed 10$\times$ under interference.  \systemname's
median remains essentially unchanged, demonstrating that the typical
request is also protected, not only tail percentiles.

\subsection{Mean Latency}
\label{sec:app-mean}

\begin{figure*}[!t]
    \centering
    \raisebox{16mm}{\rotatebox{90}{\small Mean TTFT (ms)}}%
    \begin{subfigure}[b]{0.240\textwidth}
        \centering
        \includegraphics[width=\linewidth]{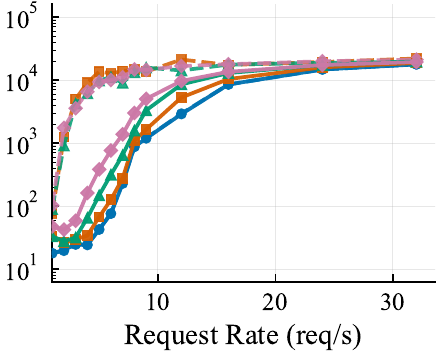}
        \caption{Llama-3 8B}
    \end{subfigure}\hfill
    \begin{subfigure}[b]{0.240\textwidth}
        \centering
        \includegraphics[width=\linewidth]{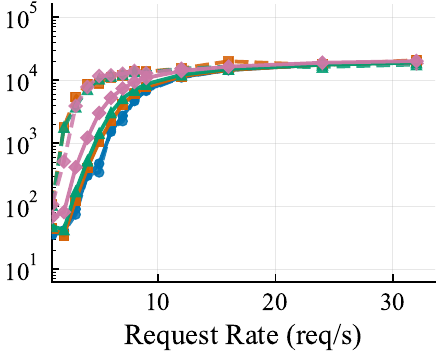}
        \caption{Phi-4 15B}
    \end{subfigure}\hfill
    \begin{subfigure}[b]{0.240\textwidth}
        \centering
        \includegraphics[width=\linewidth]{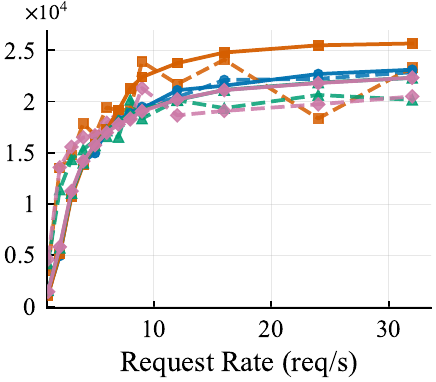}
        \caption{Qwen-3 32B}
    \end{subfigure}\hfill
    \begin{subfigure}[b]{0.240\textwidth}
        \centering
        \includegraphics[width=\linewidth]{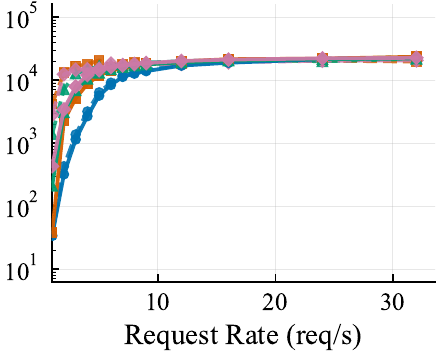}
        \caption{Qwen-3 30B-A3B}
    \end{subfigure}\\[2pt]
    \raisebox{16mm}{\rotatebox{90}{\small Mean TPOT (ms)}}%
    \begin{subfigure}[b]{0.240\textwidth}
        \centering
        \includegraphics[width=\linewidth]{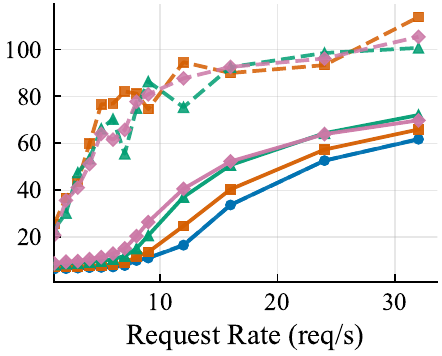}
        \caption{Llama-3 8B}
    \end{subfigure}\hfill
    \begin{subfigure}[b]{0.240\textwidth}
        \centering
        \includegraphics[width=\linewidth]{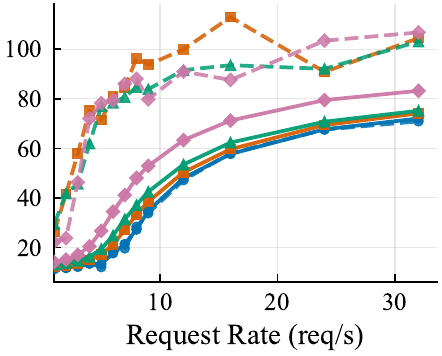}
        \caption{Phi-4 15B}
    \end{subfigure}\hfill
    \begin{subfigure}[b]{0.240\textwidth}
        \centering
        \includegraphics[width=\linewidth]{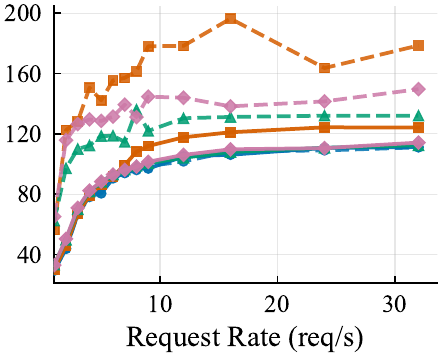}
        \caption{Qwen-3 32B}
    \end{subfigure}\hfill
    \begin{subfigure}[b]{0.240\textwidth}
        \centering
        \includegraphics[width=\linewidth]{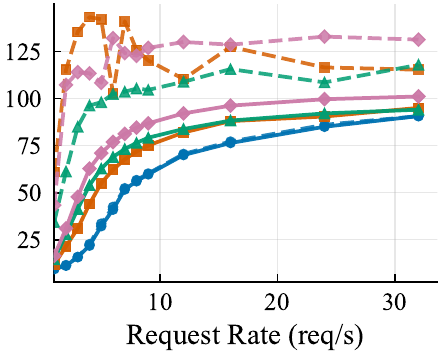}
        \caption{Qwen-3 30B-A3B}
    \end{subfigure}\\[2pt]
    \raisebox{16mm}{\rotatebox{90}{\small Mean ITL (ms)}}%
    \begin{subfigure}[b]{0.240\textwidth}
        \centering
        \includegraphics[width=\linewidth]{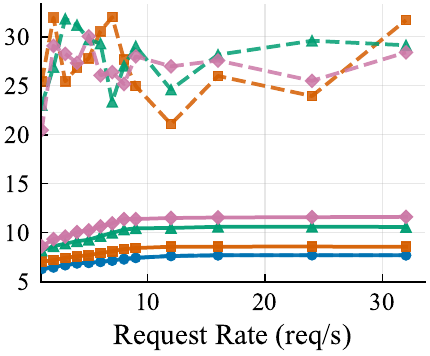}
        \caption{Llama-3 8B}
    \end{subfigure}\hfill
    \begin{subfigure}[b]{0.240\textwidth}
        \centering
        \includegraphics[width=\linewidth]{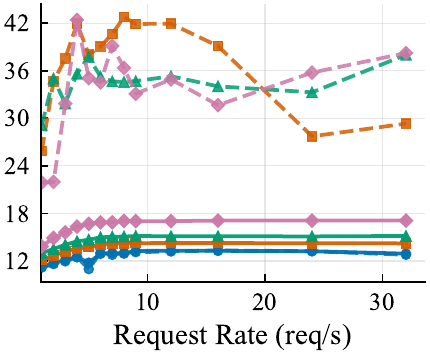}
        \caption{Phi-4 15B}
    \end{subfigure}\hfill
    \begin{subfigure}[b]{0.240\textwidth}
        \centering
        \includegraphics[width=\linewidth]{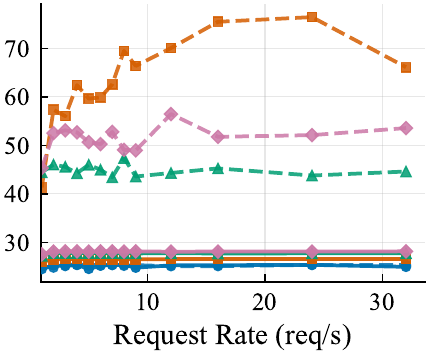}
        \caption{Qwen-3 32B}
    \end{subfigure}\hfill
    \begin{subfigure}[b]{0.240\textwidth}
        \centering
        \includegraphics[width=\linewidth]{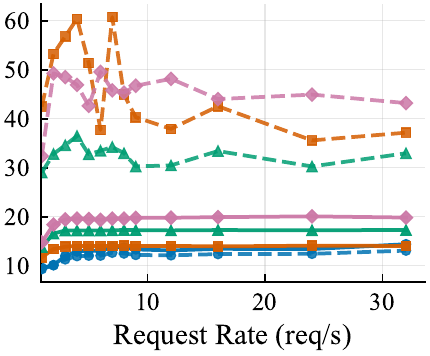}
        \caption{Qwen-3 30B-A3B}
    \end{subfigure}
    \figlegend
    \caption{Mean latency across four models.  Top row~(a--d): TTFT; middle row~(e--h): TPOT; bottom row~(i--l): ITL.  Solid lines show isolated execution; dashed lines show CPU interference.  The mean tracks between P50 and P95 due to heavy-tailed queuing distributions; system ordering remains consistent with tail percentile results.  Baseline means inflate by 3--18$\times$ under interference while \systemname's stays within 1.0--1.15$\times$.}
    \label{fig:mean-all}
    \Description[Mean latency for all models]{Mean TTFT, TPOT, and ITL across four models under isolated and interference conditions.}
\end{figure*}

Fig.~\ref{fig:mean-all} reports mean latency.  Because the mean is
sensitive to heavy-tailed distributions (common in queuing systems under
load), it tends to track between the median and P95.  The mean is
included here for readers who use it in SLO definitions or cost models.

\smartparagraph{Isolation.}
System ordering is consistent with both P50 and P99.  On Qwen-3 30B-A3B,
baselines incur 3.3$\times$ (TensorRT-LLM) to 8.1$\times$ (SGLang)
higher mean TTFT than \systemname (\qty{426}{ms})
(Table~\ref{tab:latency-summary}).  The mean captures not only the
typical case but also the impact of occasional long-tail requests, making
it a useful single-number summary for system comparison.

\smartparagraph{Interference.}
Under interference (dashed lines), mean latency shows the same pattern
documented at P99 in the main paper: baseline means inflate by
3--18$\times$ while \systemname's mean stays within
1.0--1.15$\times$ of isolation.
\systemname's steady-state path does not traverse the contended host CPU,
so even average-case latency is structurally decoupled from host load.

\section{Extended Throughput Metrics}
\label{sec:app-throughput}

The main paper reports goodput (completed requests per second).
Here we provide a complementary view: token-level throughput,
broken down into prefill (prompt tokens processed per second) and decode
(output tokens generated per second).  These metrics are relevant for
deployments that bill or provision by token count rather than by request.

\begin{figure*}[!t]
    \centering
    \raisebox{16mm}{\rotatebox{90}{\small Prefill (tok/s)}}%
    \begin{subfigure}[b]{0.240\textwidth}
        \centering
        \includegraphics[width=\linewidth]{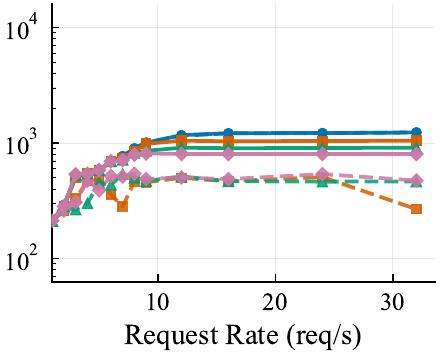}
        \caption{Llama-3 8B}
    \end{subfigure}\hfill
    \begin{subfigure}[b]{0.240\textwidth}
        \centering
        \includegraphics[width=\linewidth]{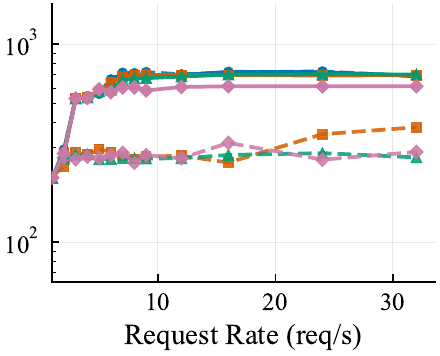}
        \caption{Phi-4 15B}
    \end{subfigure}\hfill
    \begin{subfigure}[b]{0.240\textwidth}
        \centering
        \includegraphics[width=\linewidth]{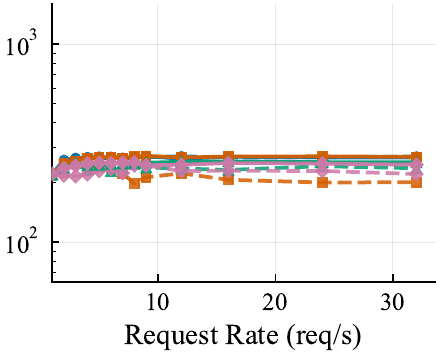}
        \caption{Qwen-3 32B}
    \end{subfigure}\hfill
    \begin{subfigure}[b]{0.240\textwidth}
        \centering
        \includegraphics[width=\linewidth]{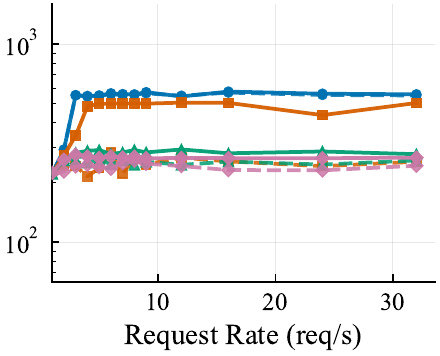}
        \caption{Qwen-3 30B-A3B}
    \end{subfigure}\\[4pt]
    \raisebox{16mm}{\rotatebox{90}{\small Decode (tok/s)}}%
    \begin{subfigure}[b]{0.240\textwidth}
        \centering
        \includegraphics[width=\linewidth]{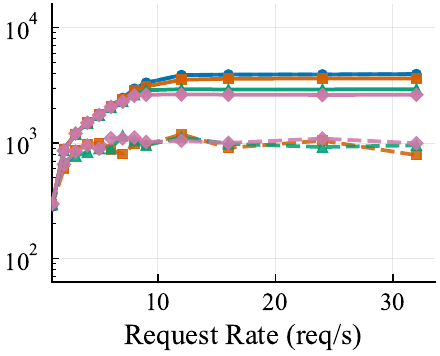}
        \caption{Llama-3 8B}
    \end{subfigure}\hfill
    \begin{subfigure}[b]{0.240\textwidth}
        \centering
        \includegraphics[width=\linewidth]{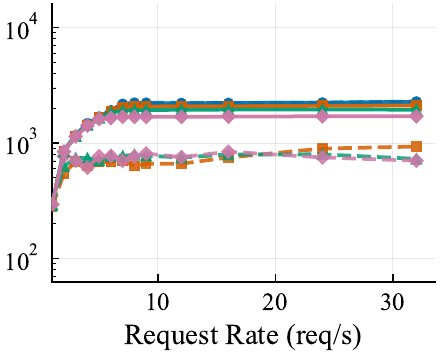}
        \caption{Phi-4 15B}
    \end{subfigure}\hfill
    \begin{subfigure}[b]{0.240\textwidth}
        \centering
        \includegraphics[width=\linewidth]{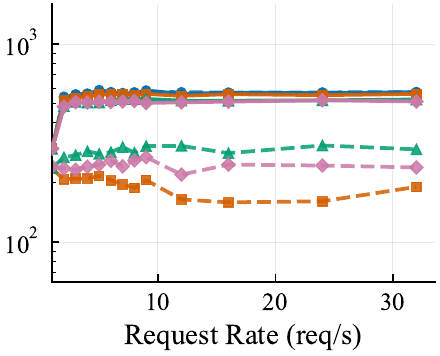}
        \caption{Qwen-3 32B}
    \end{subfigure}\hfill
    \begin{subfigure}[b]{0.240\textwidth}
        \centering
        \includegraphics[width=\linewidth]{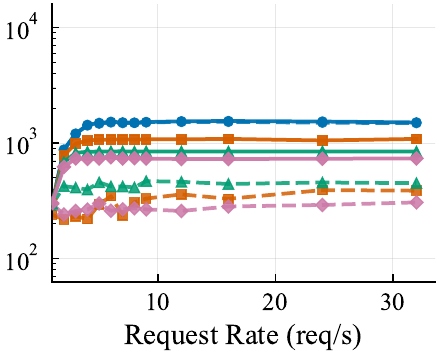}
        \caption{Qwen-3 30B-A3B}
    \end{subfigure}
    \figlegend
    \caption{Token-level throughput across four models.  Top row~(a--d): prefill throughput (prompt tokens/s); bottom row~(e--h): decode throughput (output tokens/s).  Solid lines show isolated execution; dashed lines show CPU interference.  \systemname's GPU-resident scheduling eliminates per-iteration CPU round-trips, yielding higher sustained decode throughput.  Under interference, baseline decode throughput collapses while \systemname's plateau is preserved.}
    \label{fig:throughput-tokens}
    \Description[Prefill and decode throughput for all models]{Prefill and decode token throughput across four models under isolated and interference conditions.}
\end{figure*}

Fig.~\ref{fig:throughput-tokens} shows both metrics across all four
models.  Prefill throughput reflects how quickly the system can ingest
prompts and is primarily compute-bound.  Decode throughput captures
autoregressive token generation rate and is sensitive to scheduling
overhead, since each decode step requires a scheduler iteration.

\smartparagraph{Decode throughput.}
\systemname's GPU-resident scheduler eliminates per-iteration CPU
round-trips, which translates into higher sustained decode throughput.
The advantage is most visible on Qwen-3 30B-A3B
(Table~\ref{tab:throughput-summary}), where \systemname sustains
\qty{1437}{tok/s} at saturation, 36\% above TensorRT-LLM
(\qty{1053}{tok/s}) and 97\% above SGLang (\qty{730}{tok/s}).

\smartparagraph{Prefill throughput.}
Prefill throughput differences are smaller because the prefill phase is
a single large matrix multiplication that is less affected by scheduling
overhead.  Nevertheless, \systemname maintains a consistent edge (2--18\%
above the nearest baseline) because it avoids the CPU-side latency of
marshaling prefill requests.

\smartparagraph{Interference.}
Under CPU contention, baseline decode throughput drops significantly
(dashed curves in the bottom row), while \systemname's throughput plateau
is preserved.  This mirrors the goodput findings in the main paper and
confirms that the throughput collapse under interference is not specific
to the request-level metric but extends to the token level.

\FloatBarrier

\end{document}